\documentclass{aa}
\usepackage{graphicx}
\usepackage{txfonts}
\newcommand{\be}{\begin{equation}}
\newcommand{\ee}{\end{equation}}
\newcommand{\bea}{\begin{eqnarray}}
\newcommand{\eea}{\end{eqnarray}}
\newcommand{\gsim}{\lower.8ex\hbox{$\; \buildrel > \over \sim \;$}}
\newcommand{\lsim}{\lower.8ex\hbox{$\; \buildrel < \over \sim \;$}}
\begin{document}
\title{Axially symmetric relativistic MHD simulations \\
of Pulsar Wind Nebulae in Supernova Remnants}
\subtitle{On the origin of torus and jet-like features}
\author{Luca Del Zanna\inst{1}, Elena Amato\inst{2}, 
Niccol\`o Bucciantini\inst{1}}
\offprints{L. Del Zanna \\ \email{ldz@arcetri.astro.it}} 
\institute{Dipartimento di Astronomia e Scienza dello Spazio, 
Universit\`a degli Studi di Firenze, Largo E. Fermi 2, 50125 Firenze, Italy
\and
INAF, Osservatorio Astrofisico di Arcetri, 
Largo E. Fermi 5, 50125 Firenze, Italy
}
\authorrunning{Del Zanna et al.}
\titlerunning{Axisymmetric RMHD simulations of PWNe in SNRs}
\date{Received ; Accepted}
%
%
%
%
\abstract{
The structure and the evolution of Pulsar Wind Nebulae (PWNe) are studied 
by means of two-dimensional axisymmetric relativistic magnetohydrodynamic
(RMHD) simulations. After the first imaging of the Crab Nebula with 
{\em Chandra}, a growing number of objects has been found to show in the 
X-rays spatial features such as rings and jets, that clearly cannot be
accounted for within the standard framework of one-dimensional 
semi-analytical models.
The most promising explanation suggested so far is based on the combined
effects of the latitude dependence of the pulsar wind energy flux, 
shaping the wind termination shock and naturally providing a higher 
equatorial emission, and of the wind magnetization, likely responsible 
for the jet collimation by hoop stresses downstream of the shock. 
This scenario is investigated here by following the evolution 
of a PWN interacting with the confining Supernova Remnant (SNR),
from the free expansion to the beginning of the reverberation phase.
Our results confirm the oblate shape of the wind termination shock 
and the formation of a polar jet with supersonic velocities 
($v\approx 0.5 - 0.7 c$) for high enough values of the equatorial 
wind magnetization parameter ($\sigma\gsim 0.01$).
}
\maketitle
%
%
%
%
\section{Introduction}
Pulsar Wind Nebulae (PWNe, or plerions) arise from the confinement of pulsar 
winds by the surrounding medium, usually an expanding Supernova Remnant (SNR).
The relativistic magnetized pulsar wind is slowed down to non-relativistic 
velocities at a termination shock, where the magnetic field is compressed and
the bulk energy of the outflow is converted into
heat and acceleration of particles. 
These then give rise to the synchrotron and Inverse Compton emission 
observed from plerions in a very wide range of frequencies, extending 
from radio wavelengths to X-rays and even $\gamma$-rays.

The best studied plerion is the Crab Nebula, whose emission has been
extensively investigated in all frequency bands and for which most
models have been proposed. New light on the spatial structure of the Crab
Nebula emission at high frequencies has been shed by observations made 
with the {\em Chandra} X-ray satellite (Weisskopf et al. \cite{weiss00}), 
which, thanks to the unprecedented spatial resolution, has revealed a 
number of intriguing features in the inner part of the nebula (see also 
Hester et al. \cite{hester95,hester02}).
The new details highlighted strengthen the view of the Crab Nebula as
an axisymmetric object. 
In what is thought to be the equatorial plane of the pulsar rotation,
{\em Chandra} observations show the presence of a bright ring of emission,
lying at a much closer distance to the pulsar than the already identified
X-ray torus (e.g. Hester et al. \cite{hester95}). 
The most puzzling discovery, however, is probably the presence of two 
opposite jet-like features oriented along an axis perpendicular to the 
plane of the torus and emerging from the very close vicinity of the pulsar. 
Similar features have been observed also in a number of other objects,
namely around the Vela pulsar (Helfand et al. \cite{helfand01}; 
Pavlov et al. \cite{pavlov03}), PSR 1509-58 (Gaensler et al. 
\cite{gaensler02}) and in the supernova remnants G0.9+01
(Gaensler et al. \cite{gaensler01}) and G54.1+0.3 (Lu et al. \cite{lu02}).

While the presence of a X-ray bright torus may be at least qualitatively
explained within the framework of standard 1-D RMHD models 
(Kennel \& Coroniti \cite{kc}, KC84 hereafter; 
Emmering \& Chevalier \cite{ec87}), 
if we further assume that either the energy flux emerging from 
the pulsar or the termination shock dissipation efficiency is higher at low 
latitudes around the equator, the presence of jets that seem to emanate 
directly from the pulsar poses severe theoretical problems in its 
interpretation (Lyubarsky \& Eichler \cite{lyueic}), given the 
difficulties at explaining 
self-collimation of ultra-relativistic flows. 
A recent suggestion for an answer to this puzzle 
(Bogovalov \& Khangoulian \cite{bk1,bk2};
Lyubarsky \cite{lyu02}) is that the jets are actually originating downstream
of the pulsar wind termination shock, where the flow is only mildly 
or non-relativistic. If this is the case, the fact that they are observed
starting from a much closer distance from the pulsar than where
the shock in the equatorial plane is thought to be, has to be interpreted
assuming that the given degree of anisotropy in the energy flow from the 
pulsar also causes the shock front to be highly non-spherical in shape, much 
closer to the pulsar along the rotation axis than in the equatorial plane.
Moreover, even if the pulsar wind is weakly magnetized just upstream of the
termination shock, the magnetic field inside the plerion can become as high 
as to reach equipartition. Therefore, collimation of the 
downstream flow may be easily achieved 
there by magnetic hoop stresses (Lyubarsky \cite{lyu02}; 
Khangoulian \& Bogovalov \cite{kb}), resulting in 
plasma compression toward the axis and eventually in a polar jet-like 
outflow.

Thanks to the recent progress in numerical relativistic fluid
dynamics and MHD (see Del Zanna \& Bucciantini \cite{luca1}; 
Del Zanna et al. \cite{luca2}, and references therein), we are
now able to start a more quantitative investigation of this problem by means
of computer simulations. Our aim is to clarify whether a given latitude
dependence of the pulsar wind energy flux may actually explain the 
jet-torus morphology observed at X-ray frequencies for the Crab Nebula
and other plerions, 
and, if this is the case, what are the conclusions that one may infer 
on the structure and magnetization of the unshocked pulsar wind. 
Here we present the results of a first series of long-term 2-D axisymmetric 
RMHD simulations, from which some general conclusions on the physical 
mechanisms at work and useful scalings may already be derived (see 
Amato et al. \cite{amato} for preliminary results). A similar numerical 
investigation has been recently carried out 
(Komissarov \& Lyubarsky \cite{kl}, KL03 hereafter), 
confirming the basic physical
picture as viable for explaining the main observational features, as 
strongly suggested also by the close resemblance, at least
at a qualitative level, between the map of simulated emission and 
{\em Chandra} images of the Crab Nebula. 

The paper structure is as follows. In Sect.~\ref{sec:model} the pulsar wind
model adopted at large distances from the light cylinder is sketched. 
In Sect.~\ref{sec:setup}
the numerical details of the simulations and the initial conditions
are reported. 
Sect.~\ref{sec:results} deals with the results of the simulations, split
in three sub-sections for convenience. Finally the results are summarized
in Sect.~\ref{sec:final}, where conclusions are drawn for this preliminary 
work.
%
%
%
%
\section{Pulsar wind model and pre-shock conditions}
\label{sec:model}
The key point of all attempts at interpreting the torus and the jet-like 
features observed in plerions as arising post-shock is that the energy flux
in the unshocked pulsar wind should depend on latitude as $\sin^2 \theta$
(Bogovalov \& Khangoulian \cite{bk1}; Lyubarsky \cite{lyu02}). This angular 
dependence is related to
the structure of the residual purely toroidal magnetic field, which, far 
enough from the pulsar, is expected to depend on the polar angle as 
$\sin\theta$ ({\em split monopole} models: e.g. Michel \cite{michel73}; 
Bogovalov \cite{bog99}).
We further assume that along the way to the termination shock the
residual Poynting energy flux is almost entirely converted into particle
energy flux, as in classic models, preserving the overall angular dependence.
We do not address here the fundamental problem of how the acceleration of
the outflow and the conversion of magnetic to particle energy 
may occur, the so-called $\sigma$ paradox: see however 
Vlahakis (\cite{vlahakis04}) and references therein for 
self-similar 2-D relativistic
MHD models, though many other mechanisms involving waves, reconnection
or kinetic plasma processes have been proposed 
(see e.g. Arons \cite{arons03} for a review and references therein).
Therefore, in the present work we will consider as an initial condition
an axisymmetric cold ($p\ll\rho c^2$) relativistically expanding pulsar wind 
($v_r\approx c$), 
with a small ratio between electromagnetic and particle energy fluxes 
just upstream of the termination shock, thus extending the standard
KC84 picture to 2-D. 

Following the previous analytical studies cited above, let us then introduce 
the latitude dependence of the energy flux from the pulsar as a dependence 
on $\theta$ of the wind Lorentz factor $\gamma$, namely:
\be
\label{eq:lorentz}
\gamma (\theta)=\gamma_0 [\alpha+(1-\alpha) \sin^2 \theta ],
\ee
where the subscript $0$ indicates quantities in the equatorial plane,
and $\alpha\leq 1$ is a parameter controlling the ratio between the
Lorentz factor at the pole and that at the equator. 
We then assume the streamlines to be radial upstream of the shock and the
mass flux to be isotropic (Bogovalov \& Khangoulian \cite{bk1}):
\be
\label{eq:dens}
\rho (r,\theta)=\rho_0 \left({r_0 \over r}\right)^2
{\gamma_0 \over \gamma (\theta)},
\ee
with $r_0$ the (arbitrary) distance from the pulsar on the equatorial 
plane at which the value $\rho_0$ is assigned. As it was shown in
the above cited work, these assumptions, in the case
of a weakly magnetized outflow, naturally give rise to an oblate shape of the 
termination shock front, and imply the existence, around the axis, of a 
colder, denser sector of the downstream plasma.

As anticipated, the toroidal magnetic field $B\equiv B_\phi$ is defined as
\be
\label{eq:torfield}
B(r,\theta)=B_0 {r_0 \over r}\sin \theta.
\ee
It is important to notice that the adopted pulsar wind model leads
to a ratio between Poynting and kinetic energy fluxes that does depend
on latitude but not on distance from the pulsar.
Therefore, it is convenient to take its equatorial (maximum) value $\sigma$ 
as the independent parameter controlling the wind magnetization
(thus the same quantity used by KC84) 
and let $B_0$ derive 
from it as $B_0=(4\pi\rho_0c^2\gamma_0^2\sigma)^{1/2}$. The total wind 
energy flux $F\simeq c(\rho c^2\gamma^2+B^2/4\pi)$ may then be written as
\be
\label{eq:totflux}
F(r,\theta)=\rho_0 \,c^3 \gamma_0^2 \left({r_0 \over r}\right)^2
[\alpha+(1-\alpha+\sigma) \sin^2 \theta],
\ee
and the wind anisotropy in energy flux is, at any distance, given by
the value $F(r,0)/F(r,\pi/2)=\alpha/(1+\sigma)$. 
Finally, the value of $\rho_0$ is determined
from the total pulsar spin-down luminosity,
supposed to be constant in time for simplicity (see Bucciantini et al. 
\cite{nick1,nick2} for long-term spherically symmetric simulations in the 
case of a decaying luminosity in time).

A similar wind model and pre-shock conditions were adopted by KL03, 
though the assumed mass flux was not isotropic but had the same latitude 
dependence as the energy flux ($\gamma=\gamma_0=10$ and $F(\theta)
\sim\sin\theta^2$), and the toroidal magnetic field a different shape: 
while in our case we take a field with its maximum at the equator, allowing
direct comparison with standard 1-D models and the usual definition
of $\sigma$, in the cited work the field reaches the maximum at intermediate
latitudes and then goes smoothly to zero at the equator 
(see Sect.~\ref{sec:magfield} for a detailed discussion and comparison). 
In spite of these differences, the overall latitude dependence of the 
wind energy flux, which is the quantity that really matters in shaping 
the termination shock, is in both cases the one predicted in the cited 
analytical studies.
%
%
%
%
\section{Simulation setup}
\label{sec:setup}
\subsection{Numerical settings}
The problem of the interaction of the pulsar wind sketched above with
the surrounding medium is here addressed numerically by performing
2-D axisymmetric simulations. The numerical tool employed is the 
shock-capturing code for relativistic MHD developed by Del Zanna et al. 
(\cite{luca1,luca2}).
The scheme is particularly simple and efficient, since complex
Riemann solvers based on characteristic waves are avoided in favor
of central-type component-wise techniques: the solver is defined by the two 
fastest local magnetosonic speeds and spatial reconstruction at cell 
boundaries is achieved by using ENO-type interpolating polynomials.
However, due to the extremely small time-steps required in the
present simulations and because of the high Lorentz factors involved
(numerical oscillations are most dangerous in the ultra-relativistic
regime), third order reconstruction is avoided and simpler second order
limited reconstruction is employed. Time integration is achieved by means 
of a two-stage Runge-Kutta TVD algorithm, using a CFL number of $0.5$.

The spatial grid is defined by spherical coordinates with 400 cells in the
radial direction and 100 cells in the polar angle $\theta$. 
The physical domain of the simulation ranges in radius from 
$r_\mathrm{min}=0.05$ to $r_\mathrm{max}=20$ light-years
(we assume a unit length $r_0=1$ light-year, from now on all lengths will
be expressed in light-years, time intervals in years and velocities in
units of $c$, if not explicitly stated otherwise), and a logarithmic 
stretching ($dr\sim r$) is imposed to better resolve the inner region.
Note that, in order to resolve within a few computational cells the
contact discontinuity between the lighter relativistic plasma and
the heavy SNR ejecta (typical jumps of order $10^6$, see below), artificial
compression is required, especially at large radii where resolution is
necessarily lower. This may in principle amplify spurious noise
above the threshold of dissipation by numerical viscosity, thus producing
unphysical results or even code crashing. We have verified that the 
chosen grid and scheme settings are a good compromise between resolution 
and stability.

Stationary input for all quantities is imposed at 
$r_\mathrm{min}$, where the super-Alfv\'enic wind is blowing from, 
while zeroth order extrapolation is assumed at the outer boundary.
The domain in $\theta$ is the first quadrant $(0,\pi/2)$, with reflecting
conditions for $v_\theta$ and $B$ at the polar axis to enforce axisymmetry,
and on $v_\theta$ alone on the equatorial plane. Note that all quantities
are defined in our code at cell centers, which never coincide with
the symmetry axis $\theta=0$ where singularities may occur. 
By doing so, the ghost cells technique is well suited for all
non-cartesian problems and we never find carbuncle-like effects
(see the jet propagation tests in cylindrical geometry in
Del Zanna \& Bucciantini \cite{luca1}; Del Zanna et al. \cite{luca2}).

Notice that under these particular settings, only 5 (out of 8) RMHD variables
need to be evolved in time. Moreover, the magnetic field is forced to be
in the azimuthal direction alone, and thus always perpendicular to the 
velocity vector, by the assumed geometrical symmetries.
In this case, all the specific methods to treat the divergence-free
constraint are obviously of no use, and the magnetic field may be evolved
as an ordinary fluid variable. Other important simplifications occur
in the algorithm for calculating the local magnetosonic speeds and
in that for deriving the primitive variables from the set of conservative
variables (see Del Zanna et al. \cite{luca2}). 
In order to speed up calculations, the
pre-shock quantities are not updated in time and the global time-step is 
defined at the inner termination shock radius (on the axis), which is
moving outward thus having the effect of accelerating the simulation.

\subsection{Initial conditions and simulation parameters}
The initial conditions are given as follows. The pulsar wind, modeled as
in Sect.~\ref{sec:model}, is set up within an arbitrary radius of $0.2$.
Once the two free parameters $\alpha$ and $\sigma$ are provided, the
wind is uniquely determined by the equatorial Lorentz factor $\gamma_0$
and total wind luminosity $L_0$. Appropriate values of the above parameters 
come by fits of the high-energy emission (for the Crab Nebula) based on 
spherically symmetric MHD (KC84) and kinetic (Gallant \& Arons \cite{gallant}) 
models of the wind-nebula interaction, namely $\gamma_0\gsim 10^6$ 
and $L_0\simeq 5\times 10^{38}\mathrm{erg~s}^{-1}$.

From a numerical point of view, however, such a high Lorentz factor is
well beyond the capabilities of existing relativistic MHD codes.
Here we assume $\gamma_0=100$ (notice that ours are the multidimensional 
RMHD simulations with the highest Lorentz factor available in the 
literature so far), and an {\em averaged} wind luminosity of
$5\times 10^{39}\mathrm{erg~s}^{-1}$, in order to speed up the evolution.
The resulting rest mass densities will be of course unrealistically high,
since we basically have $\rho_0\gamma_0^2=\mathrm{const}$. 
We deem that this should not a problem even in multidimensional flows, 
provided $\gamma\gg 1$ in the wind region at all latitudes (as in our case).
This property has been nicely demonstrated in 1-D by KC84, where there is no
explicit dependence on $\gamma$ and $\rho$ taken separately, while all 
quantities depend on the wind luminosity and on the magnetization parameter 
$\sigma$. In the present 2-D case, even if the termination shock is not 
circular in shape (thus the wind is not always normal to it and KC84 analysis
do not apply at all latitudes), we still find that the exact value of
$\gamma_0$ does not make any difference in the nebular flow, provided
all other parameters (except obviously $\rho_0$) are kept the same.
In particular we have performed runs with lower values of the equatorial
Lorentz factor, down to $\gamma_0=20$ (here with a small wind anisotropy, 
$\alpha=0.5$, to preserve the condition $\gamma\gg 1$ at the poles too), 
and we did not find appreciable differences in the results.

In all simulations we assume an anisotropy parameter of $\alpha=0.1$, 
thus the energy flux along the polar axis will be ten times less than 
at the equator, while density will be ten times greater.
We decided not to choose smaller values of $\alpha$ since we wanted to be 
sure that during the evolution the termination shock (TS hereafter) 
would eventually move away from the inner boundary at all latitudes, 
otherwise outflow conditions are not appropriate and may lead to unphysical 
situations, typically near the polar axis.
The magnetization parameter $\sigma$ is instead varied in the different
simulations, from $0.003$ up to $0.1$.

Around the pulsar wind region, the (spherical) expansion of the cold, 
unmagnetized supernova (SN) ejecta is simulated by setting up a region of 
high constant density with a radially increasing velocity profile 
$v=v_{ej}r/r_{ej}$, appropriate for a self-similar expansion. Here we take 
$r_{ej}=r_0=1$, while $\rho_{ej}$ and $v_{ej}$ are respectively
determined by imposing
\be
M_{ej}=\int_{0}^{r_{ej}}\rho_{ej}\,4\pi r^2dr,
\ee
where we take $M_{ej}=3M_\odot=6\times 10^{33}\mathrm{g}$, and
\be
E_{ej}=\int_{0}^{r_{ej}}{1 \over 2}\rho_{ej}\, v^2\, 4\pi r^2dr,
\ee
with $E_{ej}=10^{51}\mathrm{erg}$. The velocity at the outer boundary of 
the ejecta is then
$v_{ej}\approx 7500\mathrm{~km~s}^{-1}$, corresponding to an age of the
SNR of $r_{ej}/v_{ej}\approx 40$ years, while the velocity at the contact
discontinuity between the ejecta and the relativistic material
(CD hereafter) is $v_{ej}\approx 1500\mathrm{~km~s}^{-1}$.
Farther out, between $r_{ej}$ and $r_\mathrm{max}$, ISM conditions are
imposed, that is a uniform, static, unmagnetized background with 
$\rho_{ISM}=10^{-24}\mathrm{g~cm}^{-3}$ 
$p_{ISM}\approx 10^{-12}\mathrm{dyne~cm}^{-2}$.
For similar settings see van der Swaluw et al. (\cite{vanders01}), Blondin 
et al. (\cite{blondin}), Bucciantini et al. (\cite{nick1}; \cite{nick2}).

Finally, for simplicity we adopt here a uniform value of $4/3$ for the 
adiabatic index, appropriate for a relativistic plasma. Radially symmetric
simulations of the PWN - SNR interaction with a variable adiabatic index
($5/3$ in the ejecta and ISM regions) were performed by Bucciantini et al. 
(\cite{nick1}), to which we refer for a discussion of the complications 
implied.

Note that in KL03 the ISM is absent and ejecta expanding with a velocity 
of $5000\mathrm{~km~s}^{-1}$ are set everywhere beyond the wind region.
This allows to speed up the evolution of the TS, though important
processes due to the interaction with the external ISM (namely the
reverberation phase, see below) are completely neglected.
%
%
%
%
\section{Simulation results and discussion}
\label{sec:results}

\subsection{Overall PWN structure and evolution: 
comparison with analytical models}
\label{sec:overall}

Before studying the formation of the peculiar jet-torus structure seen
in {\em Chandra} images, let us investigate the overall plerion properties
and its evolution in time, comparing the results with analytical models,
whenever possible.

\begin{figure}
\centerline{\resizebox{\hsize}{!}{\includegraphics{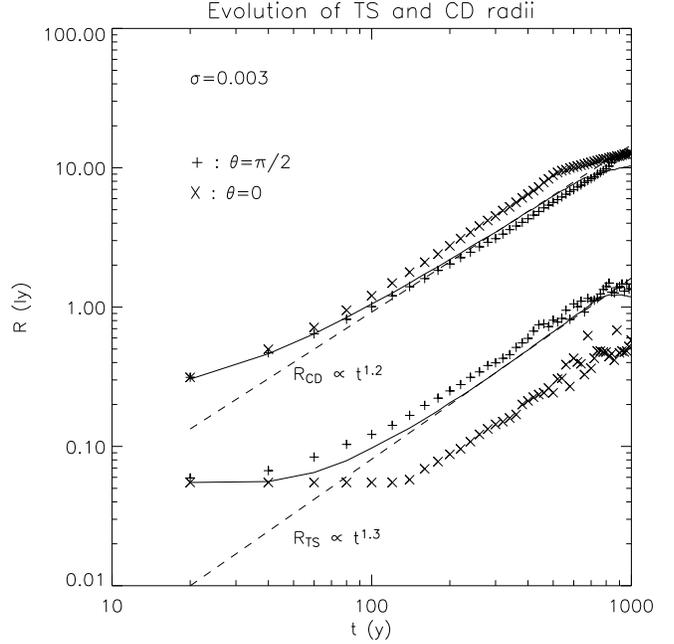}}}
\caption{
The time evolution of the PWN boundaries, the TS and CD radii,
for $\theta=\pi/2$ and $\theta=0$ (symbols as indicated on the plot), 
in the $\sigma=0.003$ case.
The 2-D results are compared with a spherically symmetric case
(solid curves)
set up with the parameters corresponding to the equatorial outflow
of the 2-D case. The self-similar 1-D hydro solution (dashed curves) is 
also shown for comparison. 
}
\label{fig:ts_cd}
\end{figure}

The PWN evolution is followed up to $t=1000$ for four cases with
different magnetization: $\sigma=0.003$, $\sigma=0.01$, $\sigma=0.03$,
and $\sigma=0.1$. 
After a short (a few years) transient stage during which, after the nebula 
is first formed, the reverse shock propagates backward, both the wind 
termination shock and the contact discontinuity (the latter separates 
the nebula from the swept up shell of ejecta) move outward. 
In Fig.~\ref{fig:ts_cd} the evolution of the PWN boundaries for 
$\theta=\pi/2$ and $\theta=0$ is plotted against time, in the
$\sigma=0.003$ case.
For a comparison, also the corresponding 1-D spherically symmetric evolution
is shown, together with the fits expected for (hydrodynamical) self-similar 
models of PWN interacting with freely expanding SN ejecta
(see Bucciantini et al. \cite{nick2}, and references therein).
At later times ($t\approx 500$ in this case) the expected self-similar 
expansion is slowed down because
of the interaction with the reverse shock produced by the motion of the
SNR in the surrounding ISM. This is the beginning of the so called 
{\em reverberation} phase (see Bucciantini et al. \cite{nick1}), here 
occurring rather early because of the high spin-down luminosity adopted.

As expected, the PWN inner boundary (the termination shock, TS hereafter) 
is farther from the pulsar at the equator than at the pole, while the 
opposite occurs at the outer boundary (the contact discontinuity, 
CD hereafter).
The former effect is due to the assumed wind energy flux anisotropy
which produces the oblate shape of the TS. The latter effect
is due, instead, to the pinching by the PWN magnetic field 
(Begelman \& Li \cite{begli}; van der Swaluw \cite{vanders03}).
Similar results are found also for more magnetized winds, 
although for higher values of
$\sigma$ the TS evolution is slower and its expansion can
begin only when 2-D effects (vortexes) remove the $\sigma<v_{CD}/c$
constraint for quasi-stationary radial MHD flows (KC84; see also
Bucciantini et al. \cite{nick2} for discussions about this constraint). 

\begin{figure}
\centerline{\resizebox{1.1\hsize}{!}{\includegraphics{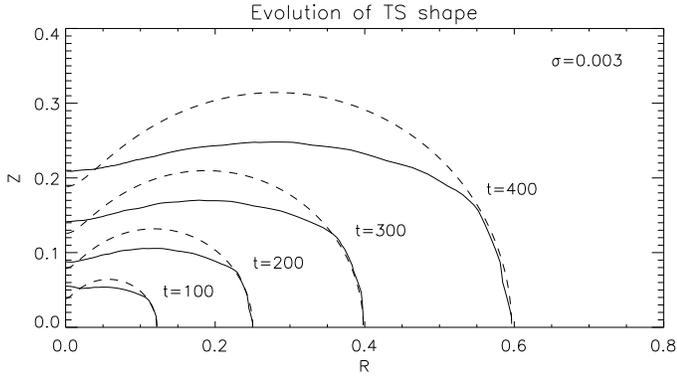}}}
\caption{
The time evolution of the TS radius in the case $\sigma=0.003$,
in the cylindrical coordinates $R$ and $Z$ (solid line). Together
with the shock shape resulting from the simulation, the expression
in Eq.~\ref{eq:ts} is plotted for comparison as a dashed curve.
Before reverberation starts ($t\approx 500$), the ratio between the polar
and equatorial radii is also well reproduced.
}
\label{fig:ts}
\end{figure}

\begin{figure}
\centerline{\resizebox{1.1\hsize}{!}{\includegraphics{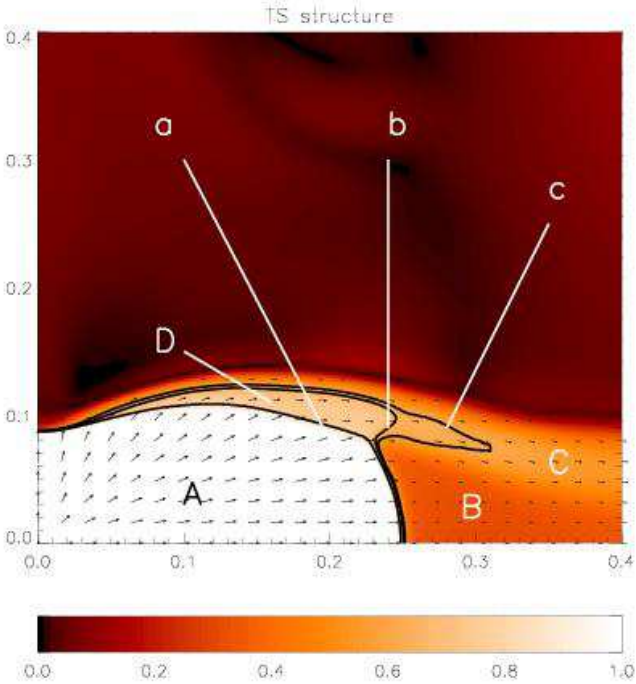}}}
\caption{
The flow structure around the TS. The background 2-D gray-scale plot 
refers to the velocity magnitude. The arrows indicate the streamlines.
Labels refer to: A) ultrarelativistic wind region; B) subsonic equatorial
outflow; C) equatorial supersonic funnel; D) super-fastmagnetosonic
shocked outflow; a) termination shock front; b) {\em rim shock}; c)
fastmagnetosonic surface.  
}
\label{fig:ts_struct}
\end{figure}

\begin{figure}
\centerline{\resizebox{1.1\hsize}{!}{\includegraphics{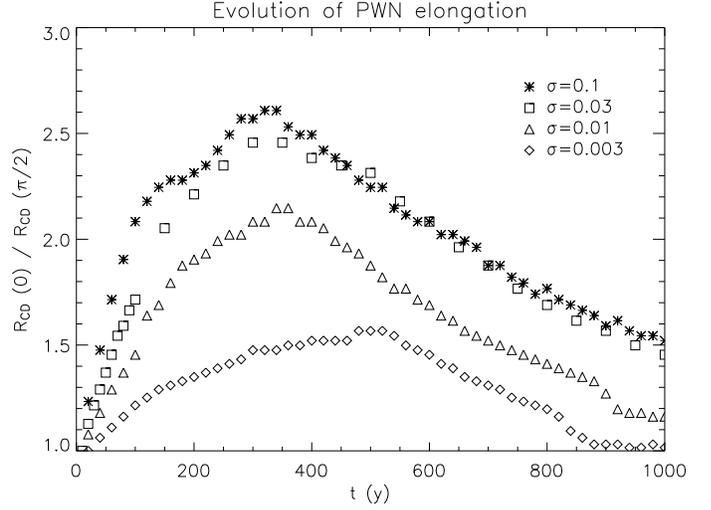}}}
\caption{
The time evolution of the PWN elongation, that is the ratio of the CD
radii at the pole and at the equator, for the various values of the wind
magnetization parameter. The elongation starts decreasing in time when 
the CD reaches the SNR-ISM reverse shock along the polar axis.
}
\label{fig:cd}
\end{figure}

Let us discuss in greater detail the above results. The time evolution of the 
TS shape is shown in Fig.~\ref{fig:ts} for the case $\sigma=0.003$.
If the downstream total pressure were constant,
the TS profile at a given time would be simply defined by the condition
\begin{equation}
F(r,\theta)\cos^2\delta=\mathrm{const},
\label{eq:tsv} 
\end{equation}
where $\delta$ is the angle
between the shock normal and the radial direction. We have verified
that the approximation of constant downstream pressure
is reasonable only within an angle $\epsilon$ around the equator, with
$\epsilon \sim 20^\circ$ for $\alpha=0.1$ as in the present
simulations. At these latitudes one also has $\delta\approx 0$,
leading to the following approximate expression for the TS profile:
\be
R_{TS}(\theta)\simeq R_{TS}(\pi/2)\left[\frac{\alpha+(1-\alpha+\sigma)
\sin^2(\theta)}{1+\sigma}\right]^{1/2}.
\label{eq:ts}
\ee
The latter expression is plotted against the numerical solution for the 
shock front in Fig.~\ref{fig:ts} for each time.
The deviations from the predicted shape at intermediate latitudes are 
mostly due to pressure variations in the post-shock region: the pressure
constancy is verified only within an order of magnitude due to the 
magnetic pinching effect and the presence of supersonic flows. 
Due to this fact, no improvement is observed if one plots the exact solution
of Eq.~\ref{eq:tsv} instead of the expression in Eq.~\ref{eq:ts}. 
Another thing to notice from the profiles shown in Fig.~\ref{fig:ts} is that
the evolution of the shock shape appears to be self-similar. We have
checked that this result holds also for all the other values of $\sigma$
considered (as long as the PWN evolution remains in the free expansion stage). 
   
The detailed structure of the flow in the vicinities of the TS
is shown in Fig.~\ref{fig:ts_struct}, where its complexity is apparent. 
Here we show the different regimes of the post-shock flow: in region D,
due to the obliquity of the TS, the speed remains super-fastmagnetosonic,
until the plasma crosses the {\em rim shock} (labeled in figure as b: 
see also KL03), and it is finally slowed down to sub-fastmagnetosonic,
yet still supersonic, speeds in the funnel C.
The TS front between the equator and the rim shock latitude $\epsilon$
(see above) is almost perpendicular to the wind (here is where Eq.~
(\ref{eq:ts}) is a good approximation), so that the latter is directly
slowed down to subsonic speeds (region B).

\begin{figure*}
\centerline{
\resizebox{1.1\hsize}{!}{
\includegraphics{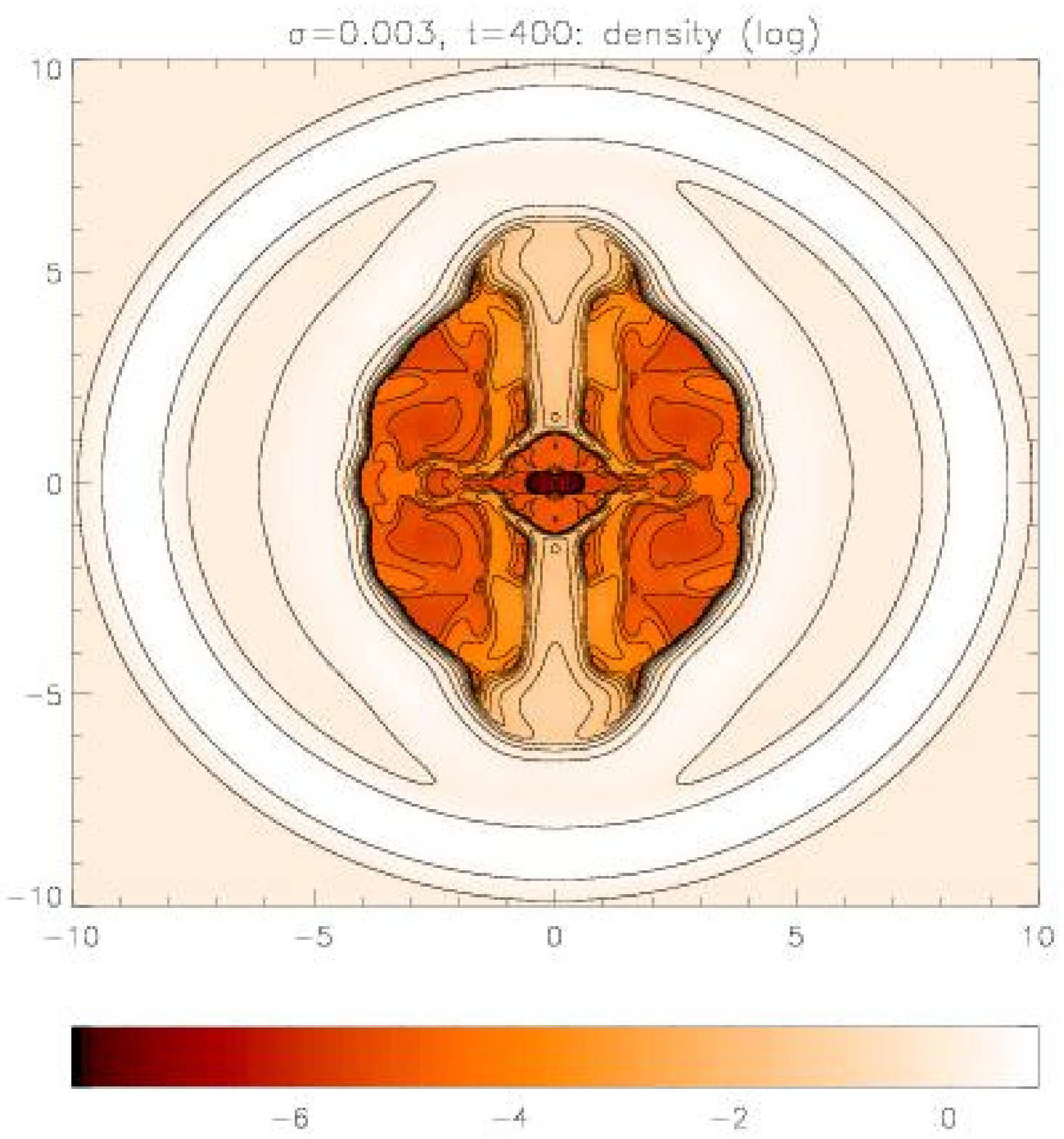}
\includegraphics{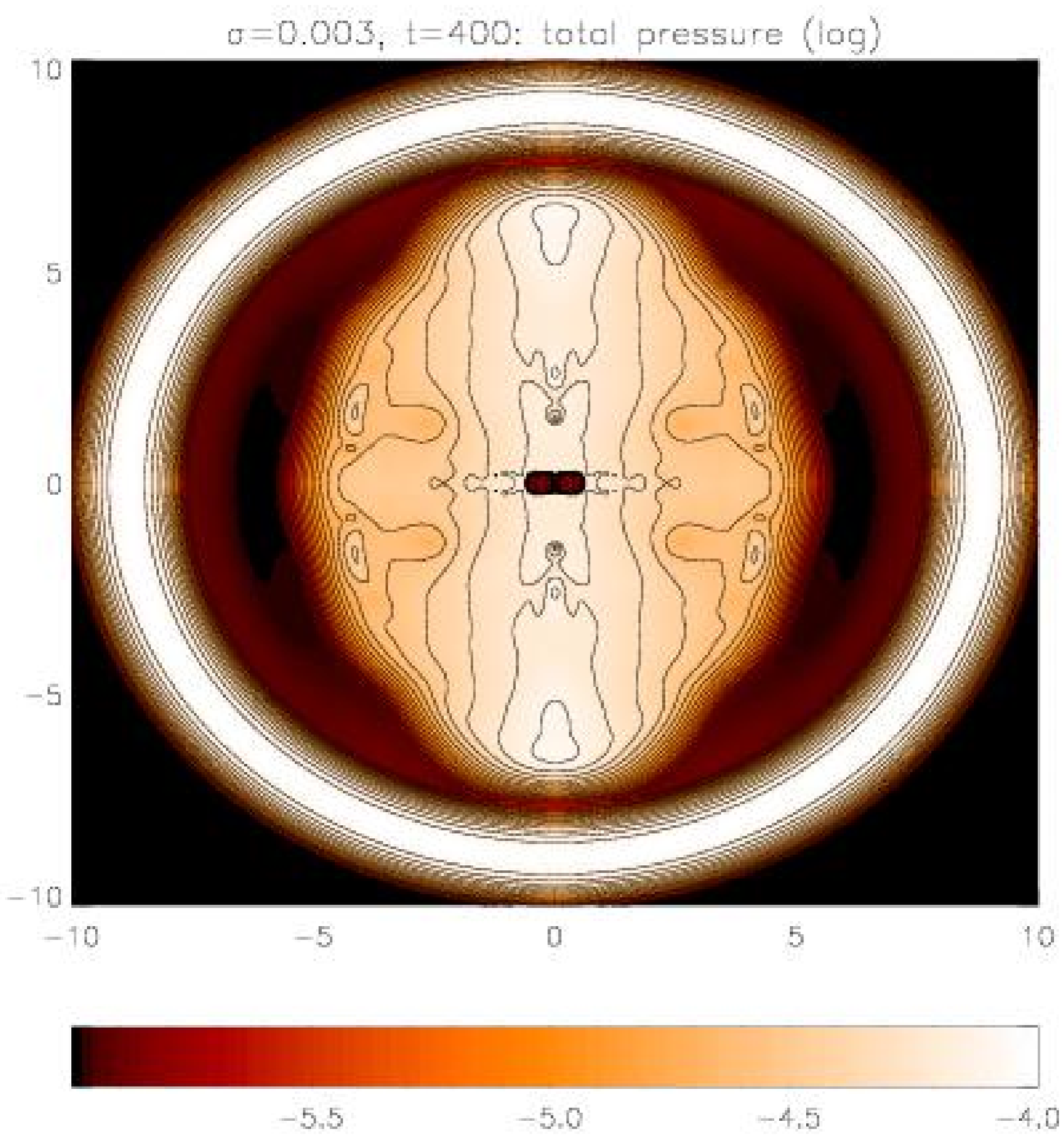}
\includegraphics{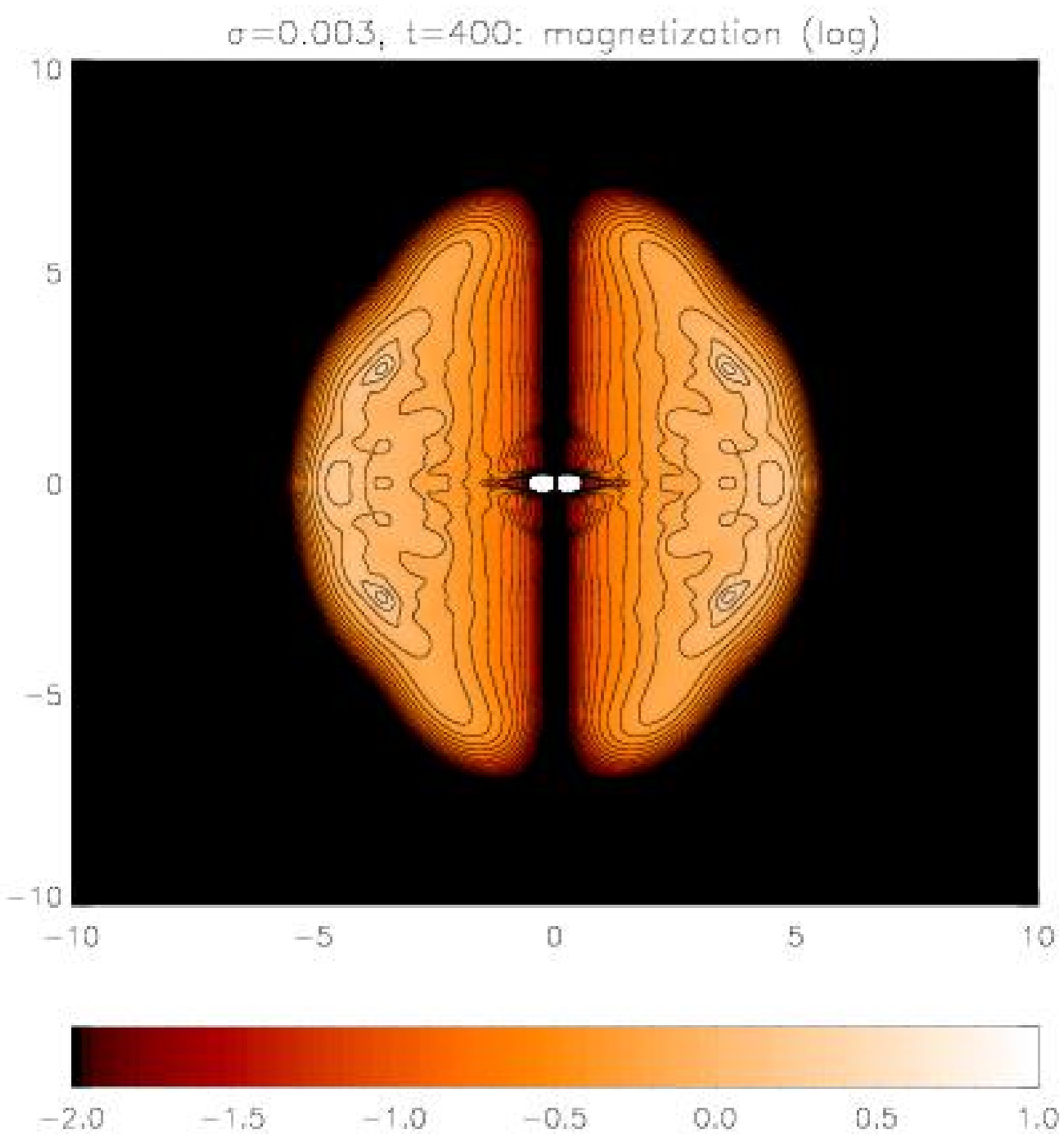}
}}
\centerline{
\resizebox{1.1\hsize}{!}{
\includegraphics{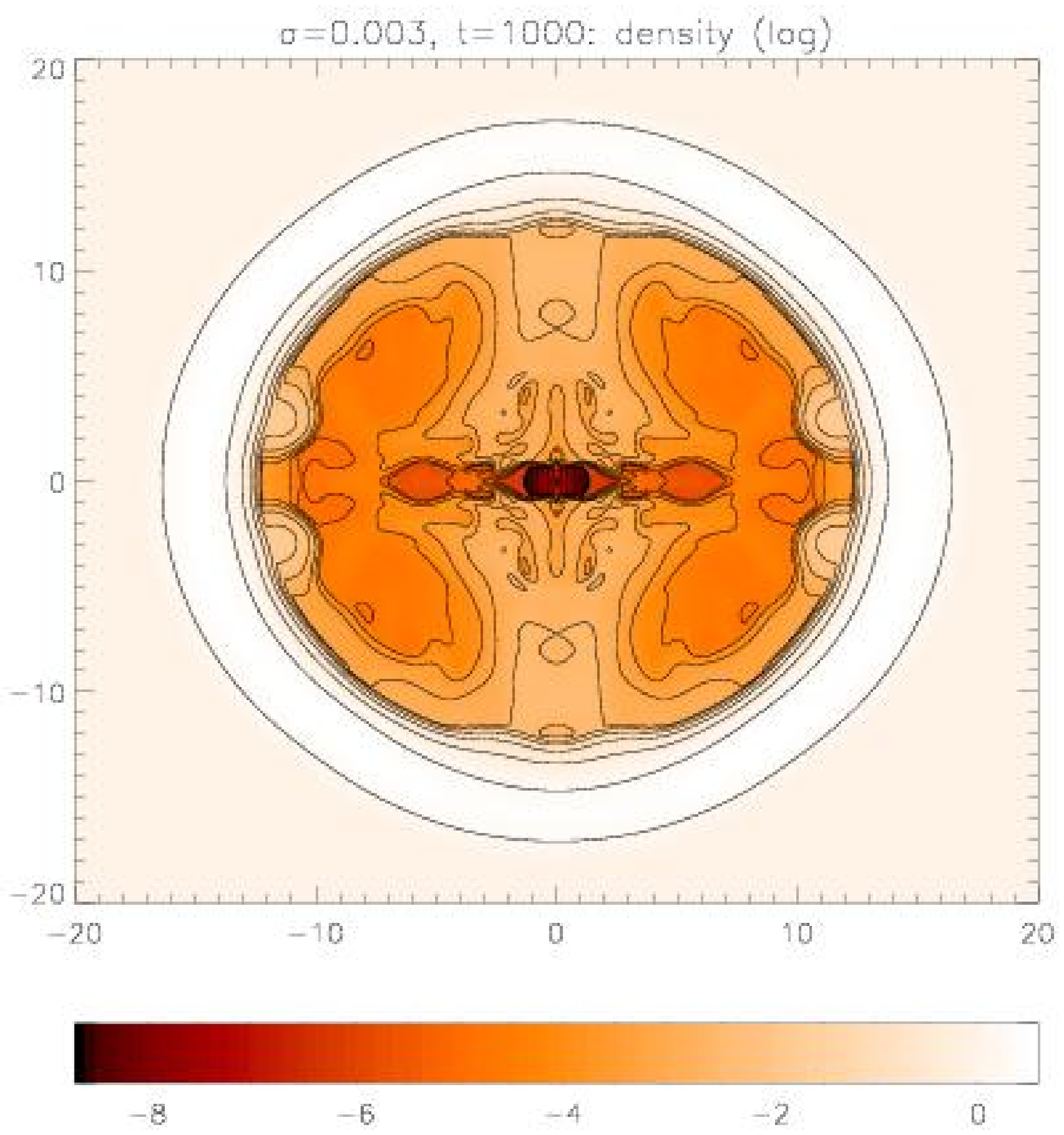}
\includegraphics{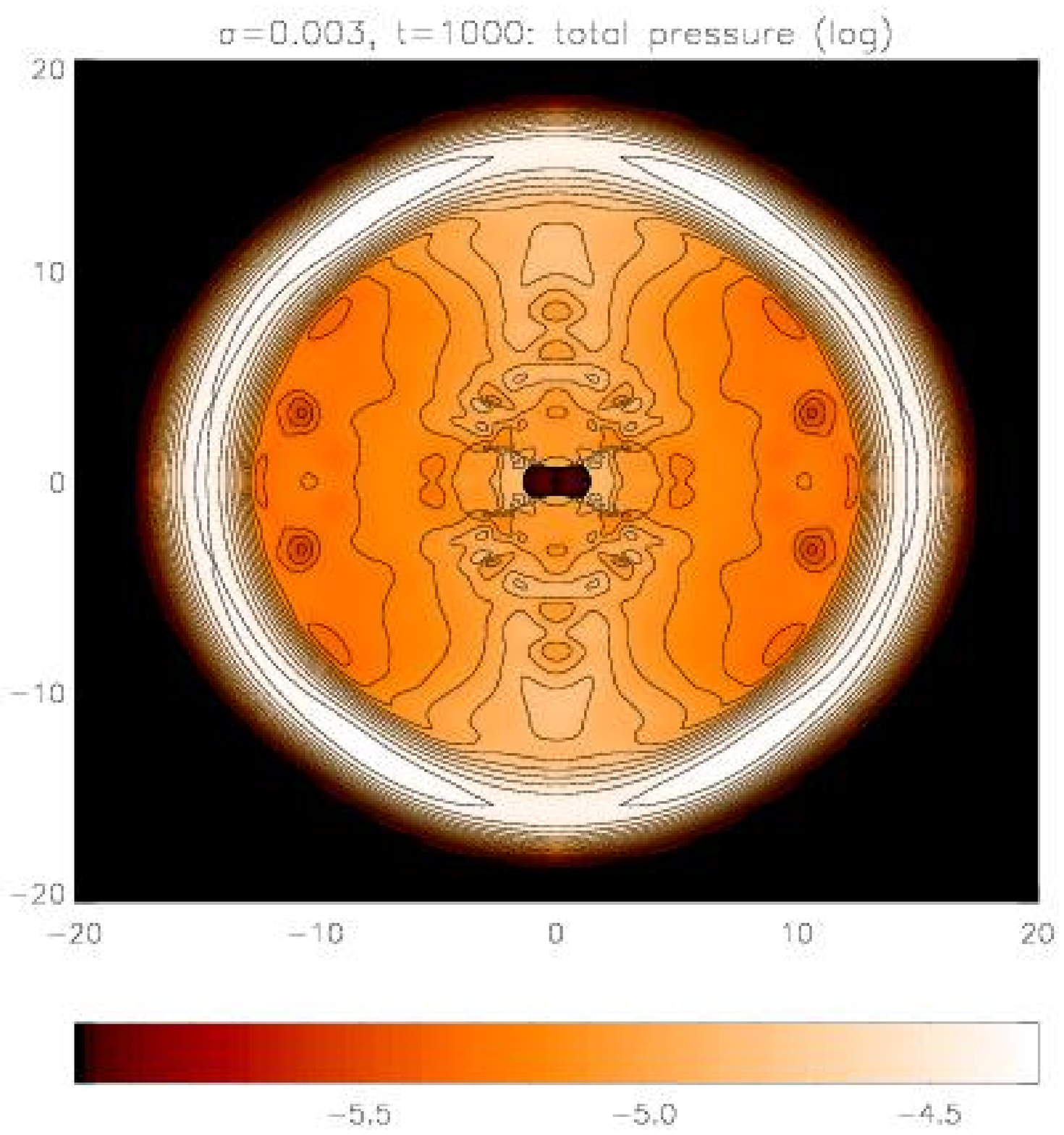}
\includegraphics{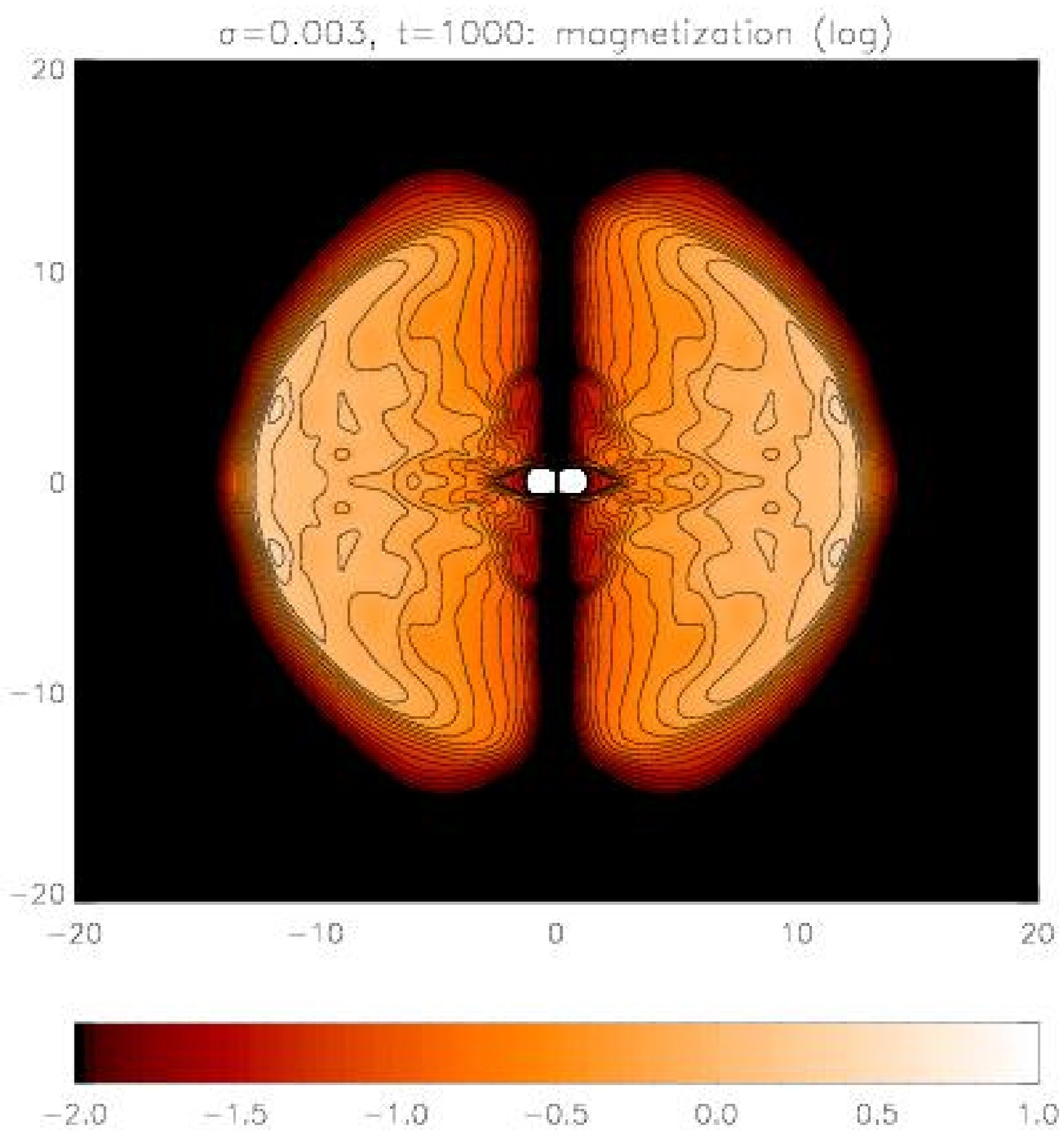}
}}
\caption{
2-D gray-scale and contour plots of the density, total pressure and
magnetization (ratio of magnetic to thermal pressure), in logarithmic 
scale. The fields are shown for
the case $\sigma=0.003$, at time $t=400$ (upper row) and $t=1000$ 
(lower row, notice that the displayed regions are different).
}
\label{fig:bl}
\end{figure*}

\begin{figure*}
\centerline{
\resizebox{1.1\hsize}{!}{
\includegraphics{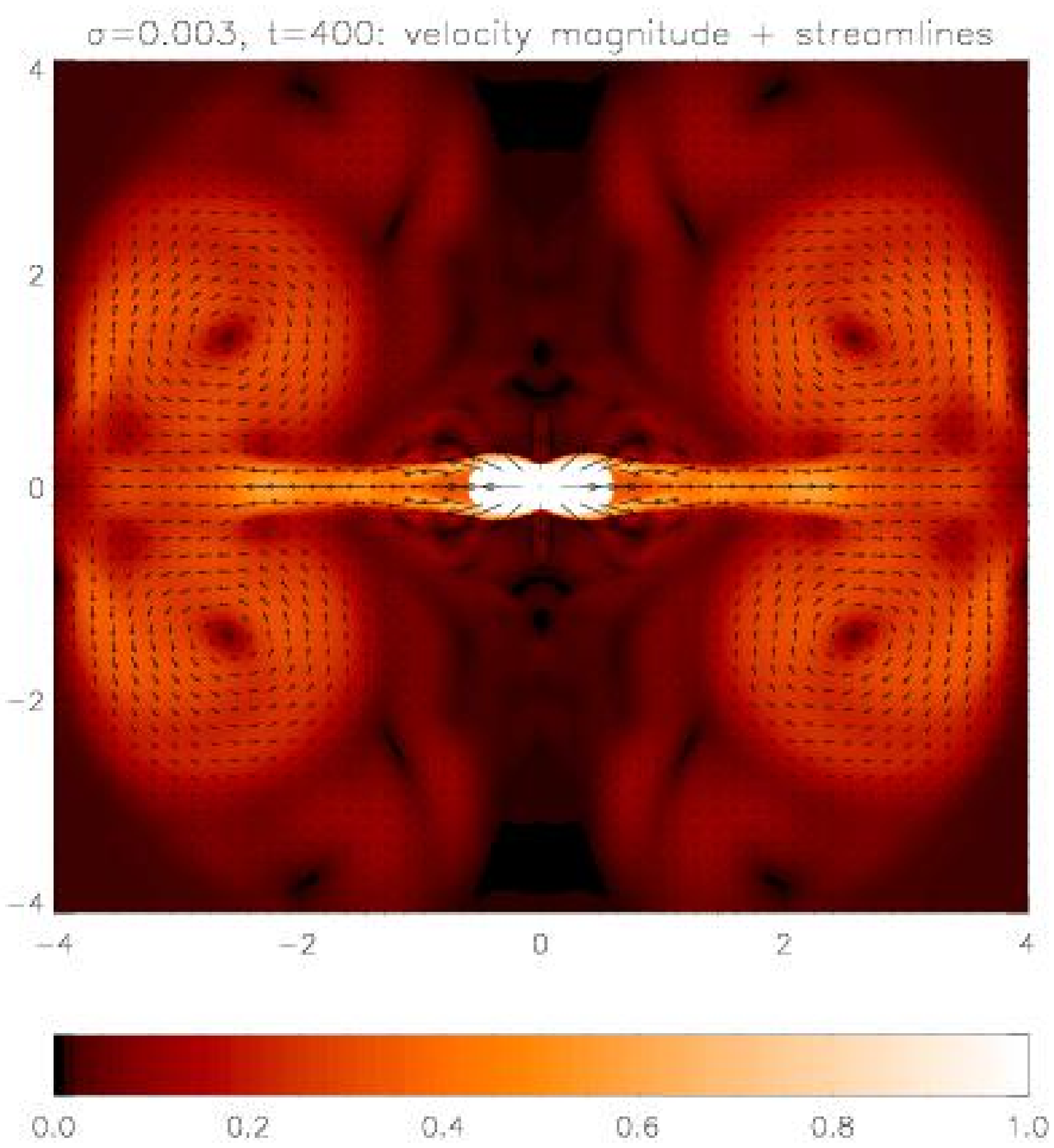}
\includegraphics{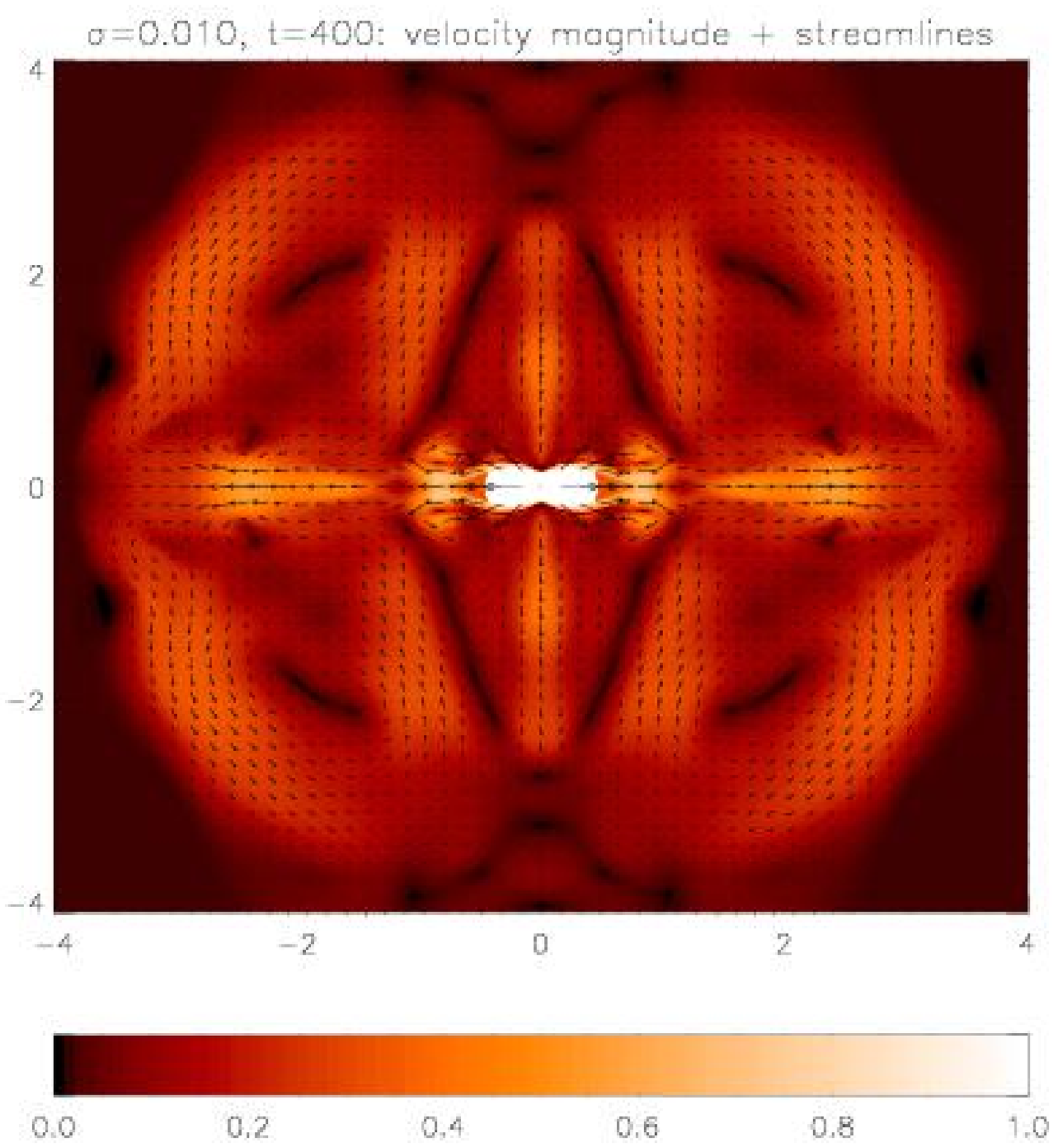}
\includegraphics{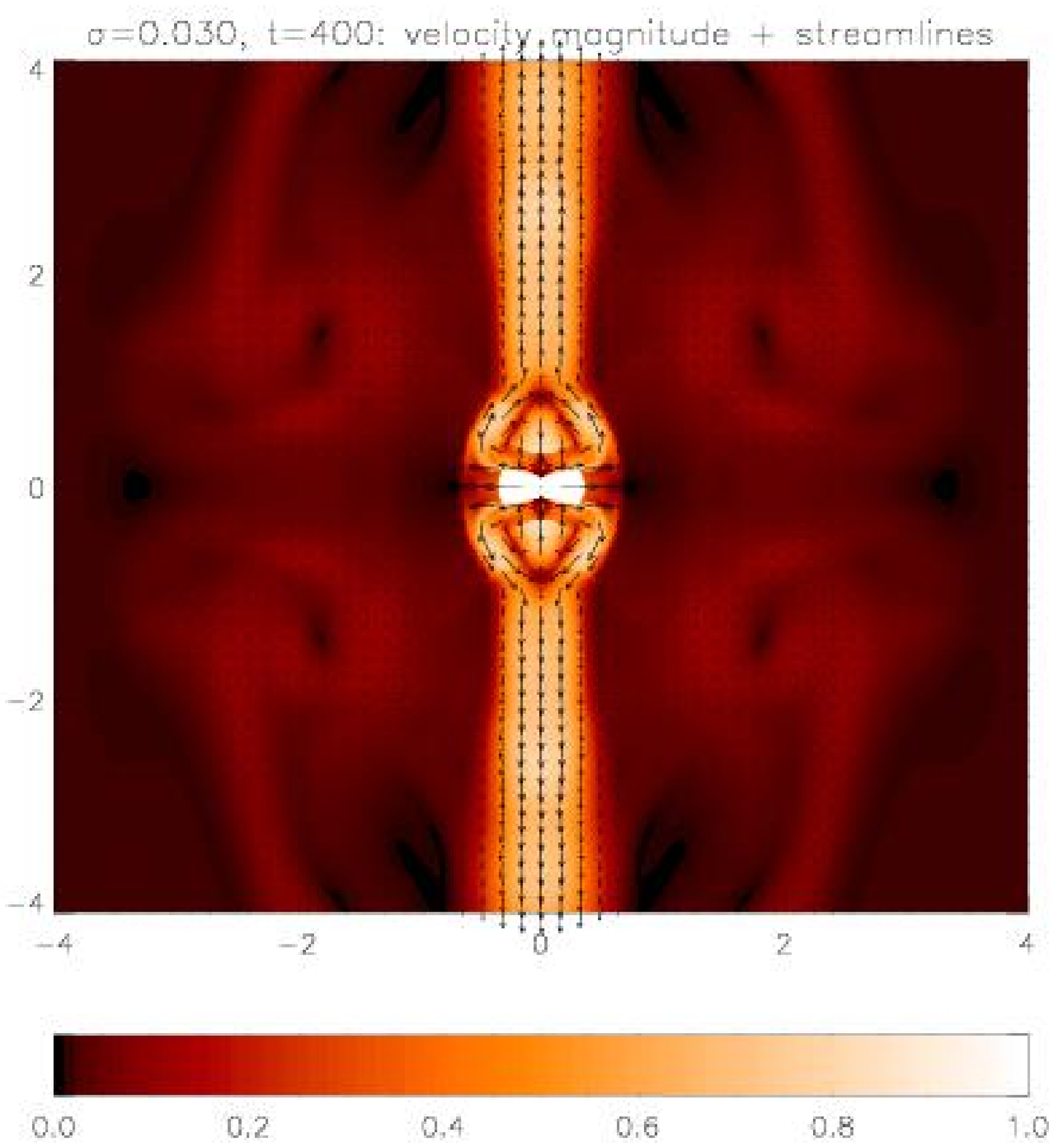}
}}
\caption{
Flow magnitude (gray scale images) and streamlines at time $t=400$
for three values of the wind magnetization parameter $\sigma$.
The jet starts to form for $\sigma=0.01$ and it is very well
developed for higher values.
}
\label{fig:jet}
\end{figure*}

\begin{figure*}
\centerline{
\resizebox{1.1\hsize}{!}{
\includegraphics{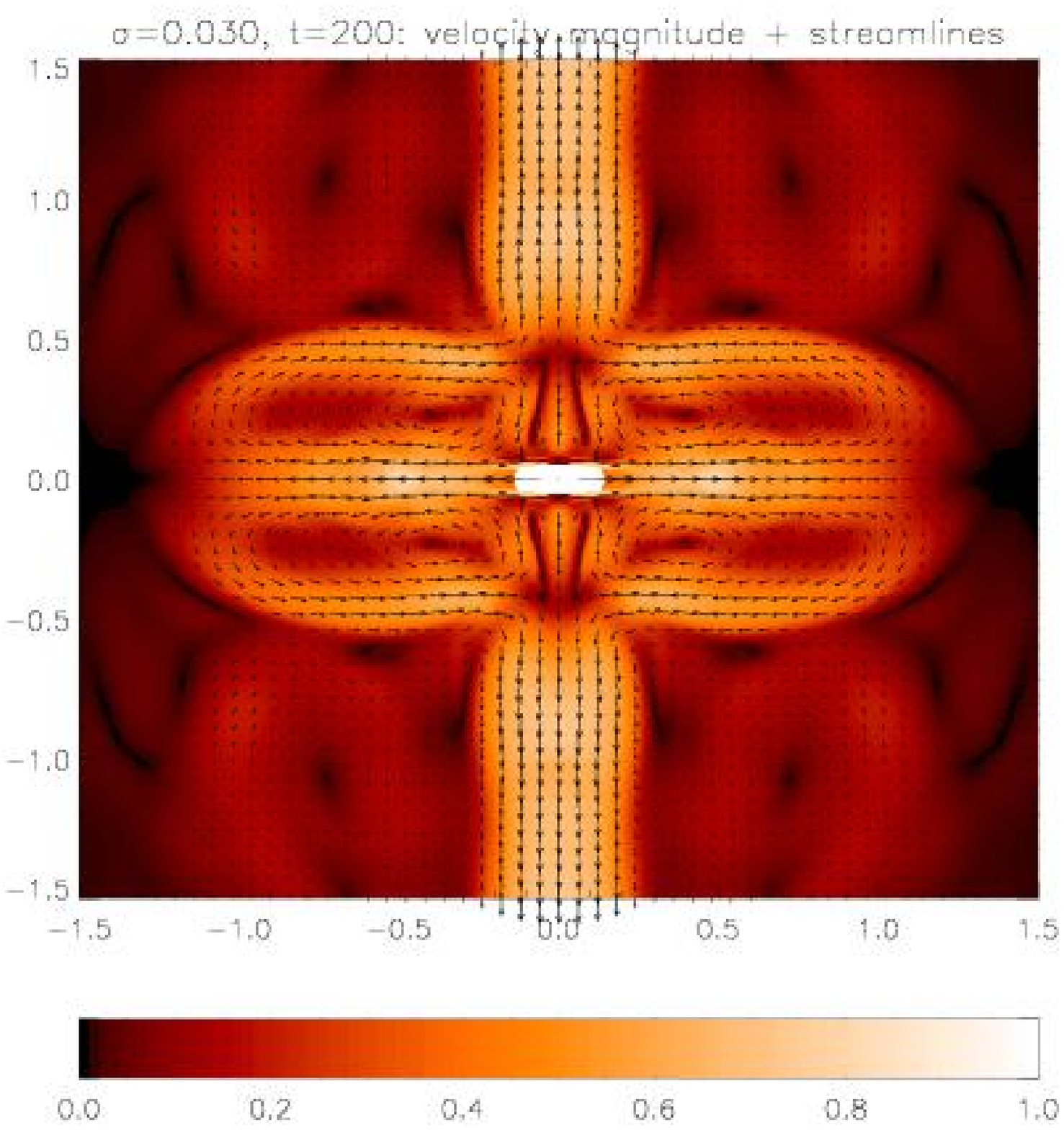}
\includegraphics{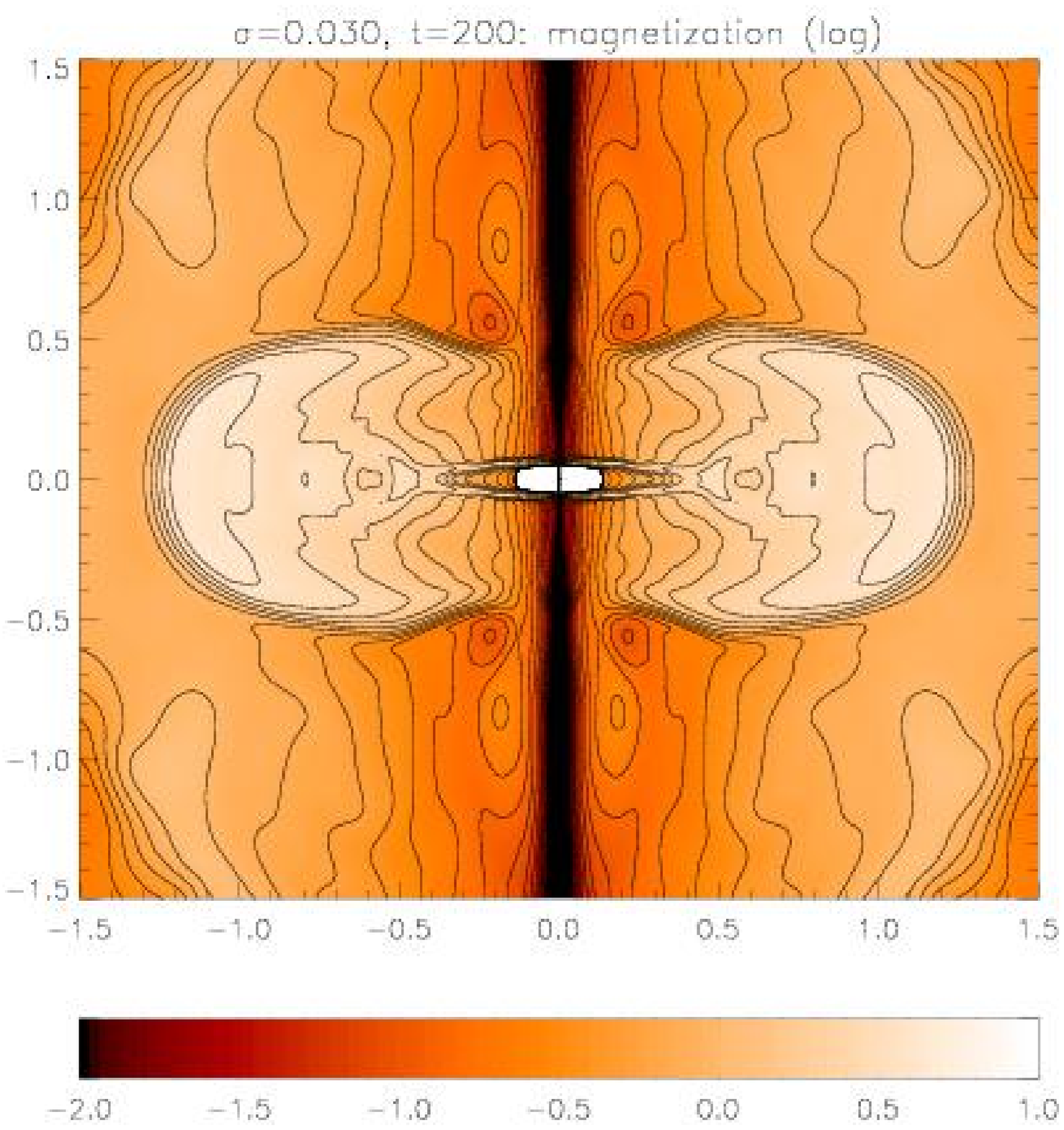}
\includegraphics{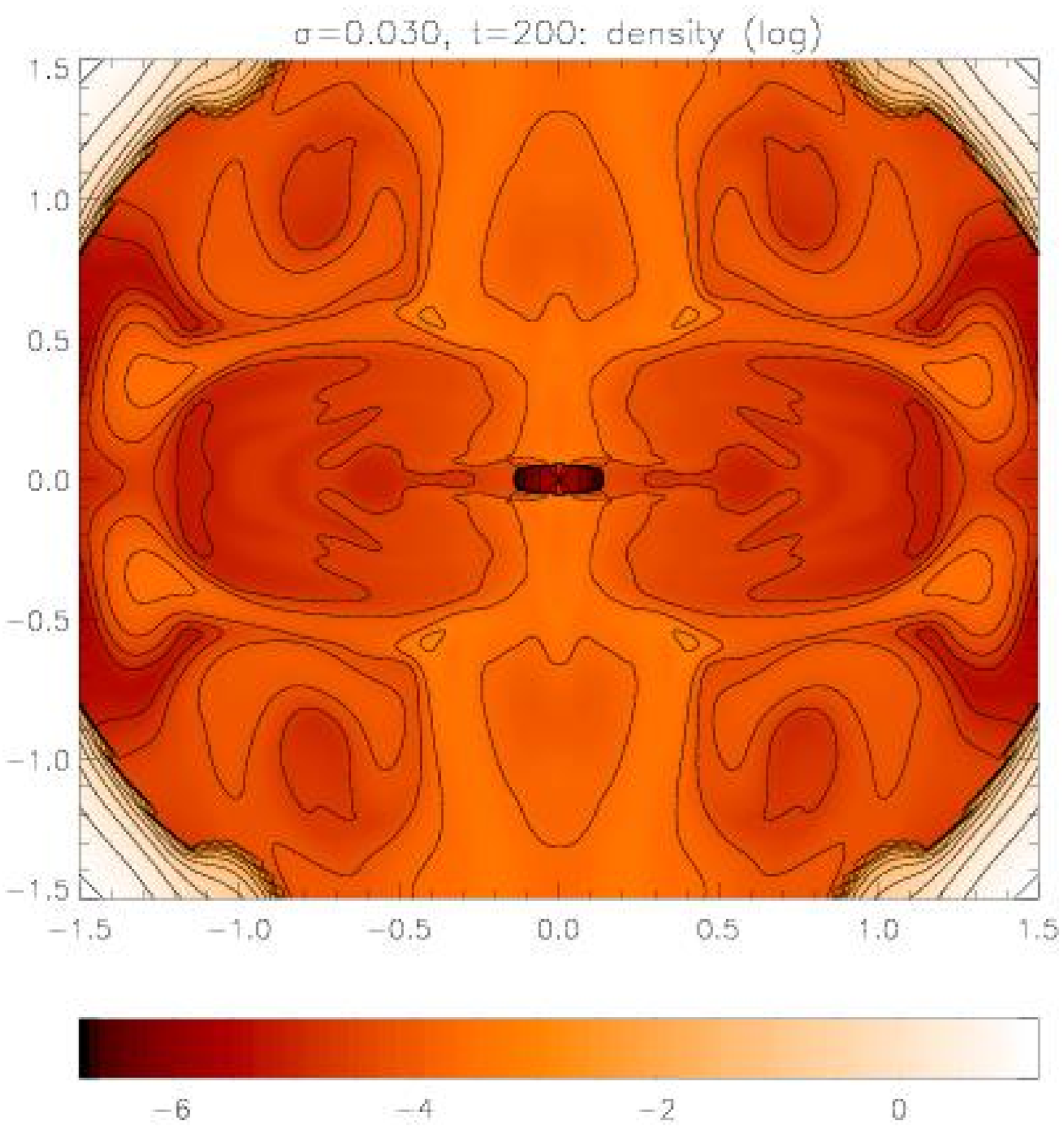}
}}
\centerline{
\resizebox{1.1\hsize}{!}{
\includegraphics{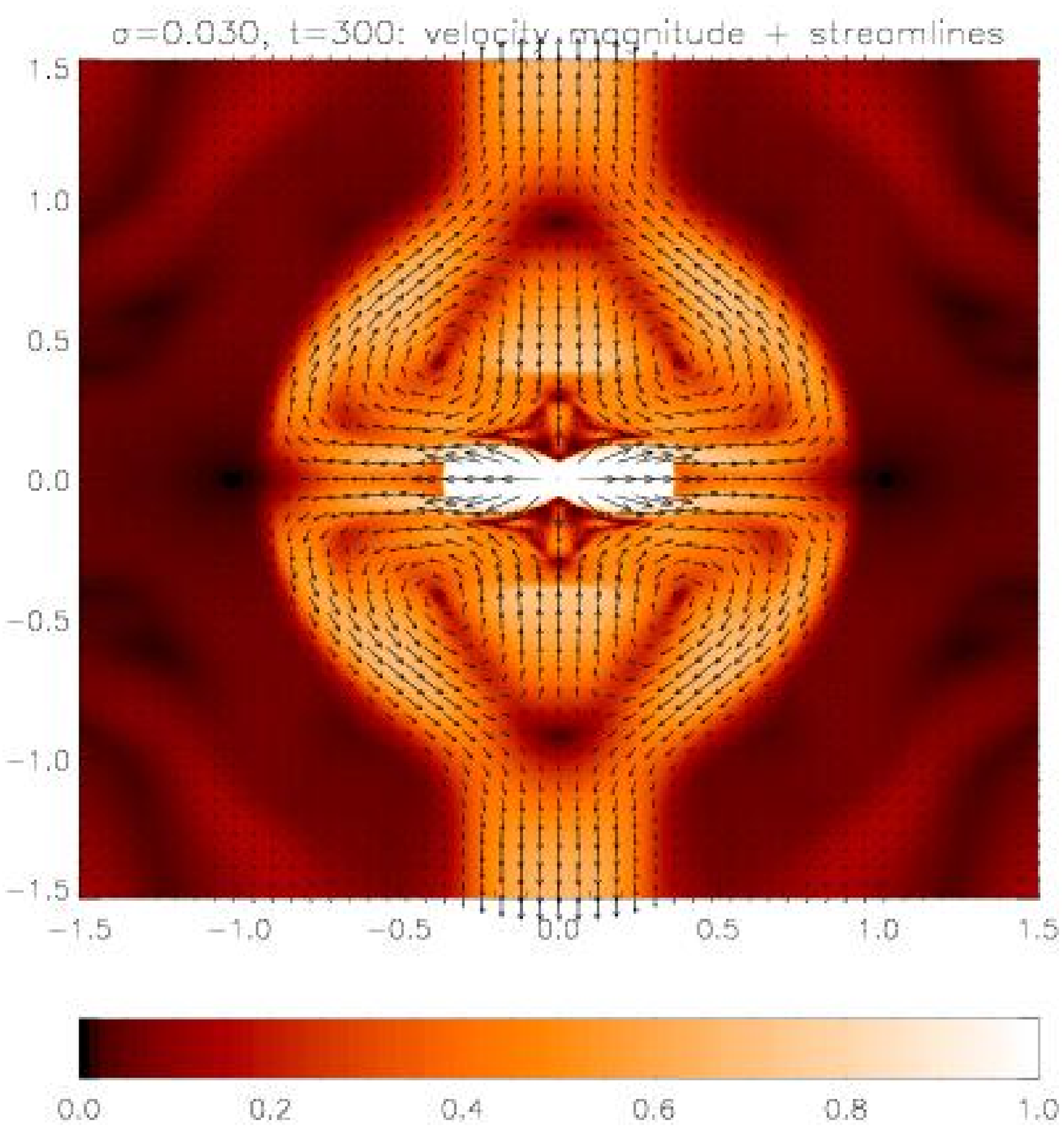}
\includegraphics{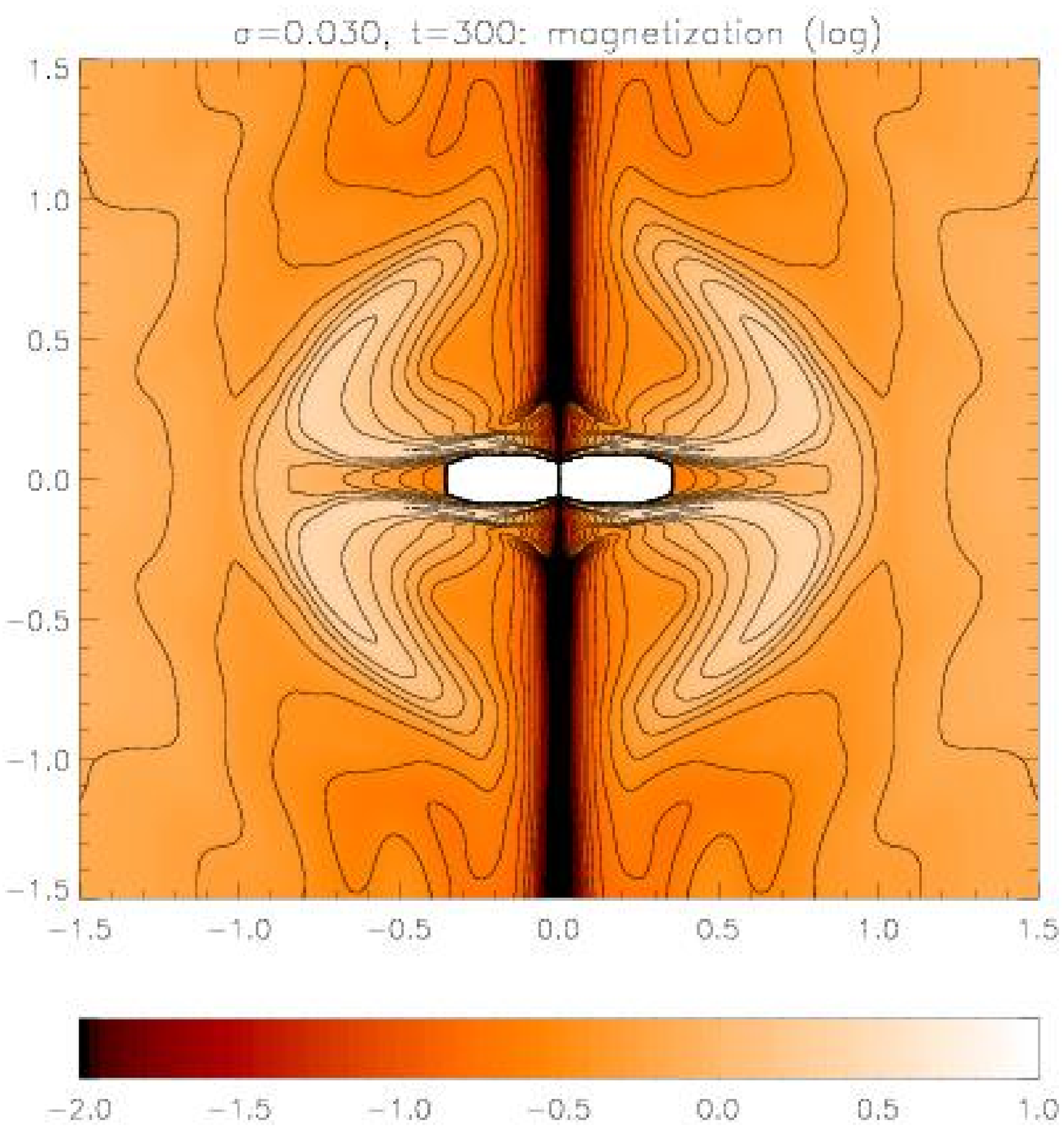}
\includegraphics{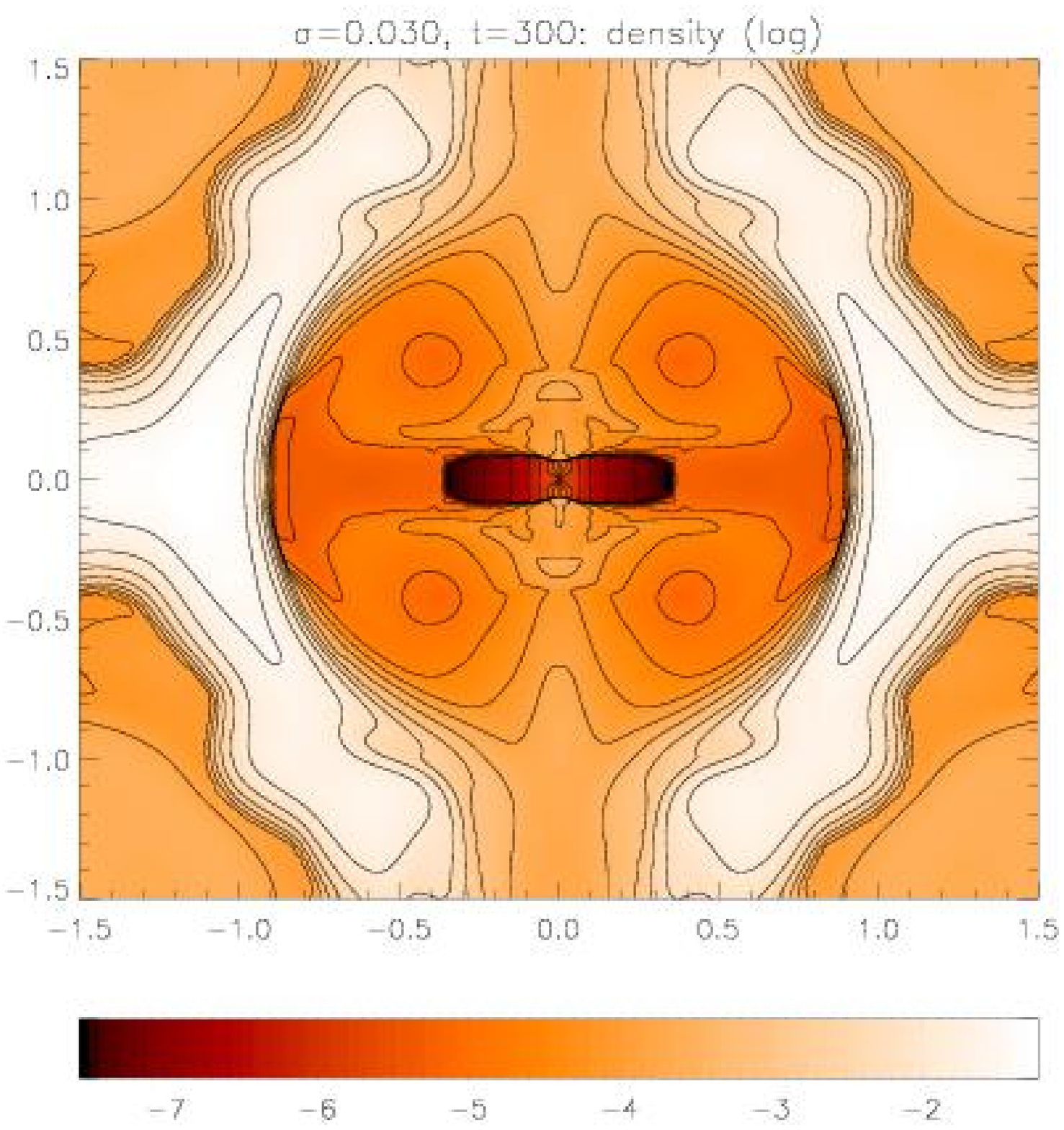}
}}
\caption{
The mechanism responsible for jet formation. Flow, magnetization
and density maps are shown for the $\sigma=0.03$ case at two different 
times: $t=200$ (upper row) and $t=300$ (lower row).
}
\label{fig:inner}
\end{figure*}

In Fig.~\ref{fig:cd} the PWN elongation, defined as $R_{CD}(0)/R_{CD}(\pi/2)$,
is shown as a function of time for all the different values of the wind 
magnetization, thus
extending the analysis by van der Swaluw (\cite{vanders03}) to the 
relativistic case
and to higher magnetizations. We confirm the result that the elongation
increases with time and with $\sigma$. The growth with time is limited
to the free expansion phase,
before the interaction with the reverse shock from the ISM.
When this interaction starts, it has two consequences: 
the reduction in time of the elongation, eventually toward unity, and the
saturation of the elongation increase with $\sigma$.
We reiterate that this interaction begins early in time in our simulations
because of the high spin-down luminosity assumed.

The situation for the low $\sigma$ case is displayed in Fig.~\ref{fig:bl},
where 2-D plots of various quantities are shown for $t=400$ and $t=1000$
(the magnetization is defined here as the ratio between the magnetic 
and thermal pressure).

At early times the PWN is clearly elongated along the symmetry axis,
since from the density and pressure plots we can see that the interaction
with the outer shell (SNR blast wave and reverse shock) has not started 
yet, except for the pole where it is just beginning.
Notice that not only the elongation, but also the expected shape
for a PWN expanding in fast moving SN ejecta is, at least qualitatively, 
reproduced (see Fig.~\ref{fig:ts}, case c, in the paper by 
Begelman \& Li \cite{begli}).
The other prediction in the cited work, namely the dependence of the total
pressure on the cylindrical radius alone inside the PWN, is reproduced
quite well, as the corresponding contours are nearly parallel to the
vertical axis. As mentioned previously, it is only in the close vicinities
of the TS that major deviations, including discontinuities, occur.
Notice, however, that such overall pressure distribution is supposed to 
hold only approximately, 
since the assumed wind energy flux is strongly anisotropic 
and vortexes of non-negligible speed may form. The presence of these
vortexes is also the cause of a dragging of high density material from
the ejecta, that starts close to the pole and later extends to the entire
nebula (see the first plot in Fig.~\ref{fig:bl}). However, this
fact does not seem to affect the overall dynamics, at least for
low $\sigma$ values. From the third
plot we can have an estimate of the relative importance of magnetic
and thermal effects, whose ratio increases with the cylindrical radius,
again as expected. 

At much later times, namely $t=1000$, the PWN has expanded so much
that the interaction with the reverse shock in the ejecta occurs at all 
latitudes. The dependence of the thermal and 
magnetic pressures on the cylindrical radius still holds approximately,
but the shape of the PWN is no more elongated, since in spite of the
pinching effect, which is still at work, the external boundaries
are now defined by the spherical SNR shell.
Due to the high value of the sound speed (actually fast magnetosonic speed)
inside the PWN, waves and disturbances created at the interaction
interface between the PWN and the SNR are rapidly transmitted and
distributed throughout the whole PWN, back to the TS that changes
in shape and slows down its evolution (see Fig.~\ref{fig:ts_cd}).

In the above discussion reference was made to the simulation
with $\alpha=0.1$ and $\sigma=0.003$ (the latter is the standard value 
deduced for the Crab Nebula from 1-D radial models, see KC84).
Let us now briefly discuss how the above results depend on these parameters.
The energy flux anisotropy is crucial for determining the TS shape,
see Fig.~\ref{fig:ts} and Eq.~(\ref{eq:ts}), though runs with different
values of $\alpha$ show that the overall morphology and evolution of
the PWN are basically unchanged.
More important is the value of the wind magnetization parameter $\sigma$, 
which determines the PWN elongation and, as a side effect, also the 
time at which the interaction with the reverse shock begins 
(see Fig.~\ref{fig:cd}).
In the following section we will discuss in greater detail the
role played by the magnetization in the physics of the PWN and in
particular in the formation of the polar jets.

\subsection{Formation of jet-like features}
\label{sec:jets}
Let us now investigate how
the flow pattern inside the PWN is affected by the nebula magnetization.
In Fig.~\ref{fig:jet} we show the speed magnitude and the streamlines
for increasing values of $\sigma$, namely 0.003, 0.01 and 0.03, at the
same time $t=400$. The jet is basically absent in the low magnetization
case, where only a subsonic flow ($v\lsim 0.1$) is observed along the
polar axis. In the intermediate case the polar outflow starts to be
more collimated and its speed increases to supersonic velocities. 
Finally, in the high magnetization case, a strong, well collimated jet
is clearly apparent. It has supersonic speed reaching values as high as
$v\approx 0.7-0.8$.

The presence of the polar jet seems to be directly correlated
to the flow pattern in the rest of the nebula. In the first
two cases an equatorial flow with speeds $v\approx 0.5$ 
is present, together with large scale vortexes at higher latitudes.
The fast equatorial flow is entirely due to 2-D effects,
since the streamlines bend toward the equator after crossing the
toroidal TS (Bogovalov \& Khangoulian \cite{bk1}). The vortexes 
occur when this equatorial
flow hits the expanding CD boundary and a circulating back-flow
is created. This pattern is present in hydro simulations as well
with the unmagnetized case being very similar to the $\sigma=0.003$ case 
in Fig.~\ref{fig:jet}. With increasing value of the wind magnetization
the equatorial outflow, for this choice of Poynting flux dependence on
latitude, is progressively suppressed. For $\sigma=0.03$ the flow
in the equatorial plane is limited to the close vicinities of the TS. 
A situation very similar to the latter is found in the highest 
magnetization case we considered,
$\sigma=0.1$, which is not displayed in the figure.

As mentioned above, a polar subsonic outflow is present even in the hydro 
simulations. The reason for this is that large scale vortexes 
eventually reach the high latitude regions near the axis, compress the 
plasma there and drive a polar flow. In addition to this process, as
the $\sigma$ of the wind increases, the magnetic hoop stresses cause
the formation of the jet-like feature with super-fast-magnetosonic 
speeds: the tension of the toroidal magnetic field, 
amplified downstream of the TS, diverts
the equatorial flow toward the polar axis, well before the large scale
vortexes due to interaction with the CD outer boundary are even formed.

This mechanism has been theoretically predicted by Lyubarsky (\cite{lyu02}),
analysed by Khangoulian \& Bogovalov (\cite{kb}) and 
finally confirmed numerically by KL03. In particular, in the 
last cited work
the authors show that significant polar velocities are found when
the effective wind magnetization parameter is $\sigma\approx 0.01$
or higher. In spite of the different settings (especially the shape
of the pulsar wind magnetic field: see next section) it is interesting 
to notice that we basically confirm here a similar value for the threshold
between cases in which the polar outflow is well developed and supersonic
and cases in which it is not. 

Let us consider the case $\sigma=0.03$ and follow more closely what
happens in the inner regions around the TS.
In Fig.~\ref{fig:inner} we plot the velocity field, magnetization and density
at two different times: $t=200$ (upper row) and $t=300$ (lower row).
By following the streamlines in the first panel we can
understand the launching mechanism. Due to the highly non-spherical shape
of the termination shock, 
narrow channels of supersonic post-shock flow form along its boundary, and
then converge toward the equatorial plane, where they merge in the high
speed ($v\approx 0.7$) equatorial beams present also in low-$\sigma$
and hydro simulations. However, this flow is no longer able to reach the
outer boundary of the PWN. After a few TS radii, the hoop stresses due
to the high magnetization in the inner part of the nebula (even beyond
equipartition: see the second plot) are so strong as to inhibit this
flow completely, thus a back-flow and vortexes form there.
When this back-flow reaches the polar axis, part of it directly escapes
along the axis, another part of it circulates back into the equatorial flow,
and yet another part heats up the plasma in the cusp region just on top
of the TS, driving again a polar outflow. The final result is then the
complete suppression of the equatorial flow at a small radius and 
the formation of a high velocity ($v\approx 0.7$) polar jet.

At $t=300$, additional effects have occurred. The TS has moved 
farther out, while the equatorial outflow region has
started shrinking, with the result of enhancing the jet driving mechanism
even more. The reason for this unexpected behavior can be appreciated
by the density maps. Fingers of cold, dense material protruding from the 
ejecta through the CD crunch the highly magnetized region of the PWN.
Rather than the result of some instability (Rayleigh-Taylor
and/or Kelvin-Helmoltz) this density enhancement coming from the external
equatorial regions is rather due to the large scale circulation induced
by the jet interaction with the slowly expanding ejecta. The polar flow
is suddenly diverted along the CD all the way down to the equatorial 
region, where it goes back toward the center eventually dragging with 
it the dense material from the ejecta (faint vortexes with $v=0.1 - 0.2$
are present also in the last plot of Fig.~\ref{fig:jet}).

While the inner nebula region is very well resolved, due to the
logarithmic radial stretching of the computational grid, one
may wonder whether the outer large scale vortexes and the dragging of
the dense material from the CD are indeed numerically converged
features, since resolution becomes correspondingly poor at large distances.
In order to check this point we have performed a long-term run with
double resolution ($800 \times 200$). The results show that the overall
structure is maintained, especially in the inner region (corresponding
to the X-ray most luminous region). However, the higher resolution allows
the development of smaller scale vortexes and instabilities, so that the
dragging is more efficient and saturation occurs earlier, although the
global behaviour is preserved.

To summarize, the pulsar wind anisotropy is responsible for the non-spherical
shape of the TS, which makes the material flow along its boundary to form
eventually the high velocity outflow in the equatorial plane.
If the wind magnetization parameter $\sigma$ is high enough, equipartition
is reached in the PWN equatorial region very near the TS and this flow
can be completely suppressed by hoop stresses and diverted toward
the polar axis, where part of it will drive the super-fastmagnetosonic jet.
When this polar outflow starts
interacting with the slowly expanding ejecta, the material escapes
along the CD down to the equator, where high density cool plasma is
dragged by the large scale vortexes toward the center causing the
crushing of the inner part of the nebula.

\subsection{Dependence on the wind magnetic field shape}
\label{sec:magfield}
When defining the wind properties we have assumed a $\sin\theta$
latitude dependence for the toroidal field, thus the magnetization
is taken to be highest at the equator. However, since the
toroidal magnetic field is the residual of the global field at
large distances from the pulsar, around the equatorial plane it is
bound to change sign. Even assuming the simplest case, that is the
split monopole model (Michel \cite{michel73}) for a perfectly aligned
magnetic rotator, at large enough distances from the light cylinder 
$r\gg c/\Omega$ ($\Omega$ is the pulsar angular velocity) the toroidal
field eventually becomes
\be
B_\phi=\frac{\Omega r\sin\theta}{v_r}B_r,
\ee
and since $B_r$ changes sign at $\theta=\pi/2$ also $B_\phi$ is bound to do so.
If the pulsar dipolar field is not exactly aligned with the
rotation axis, the so-called {\em striped wind} on the equator will
present a sequence of fieldlines of alternating sign. If magnetic
dissipation (reconnection) is efficient, the
effective toroidal field at large distances will be approximately zero
below a certain small angle (depending on the obliquity of the
rotator) around the equator (e.g. Coroniti \cite{coroniti}).

In our analysis we have considered an approximately aligned rotator, 
so that the striped wind region is so small to be dynamically
unimportant and the assumption $B\sim\sin\theta$ remains valid even
for $\theta\to\pi/2$ (the change of sign is superfluous, since $B$
appears only squared in the dynamical equations). As mentioned in 
Sect.~\ref{sec:model}, in KL03 the opposite scenario is 
considered: the striped wind region
is very large, such as to extend to intermediate latitudes affecting
the overall field shape.

\begin{figure}
\centerline{\resizebox{0.9\hsize}{!}{\includegraphics{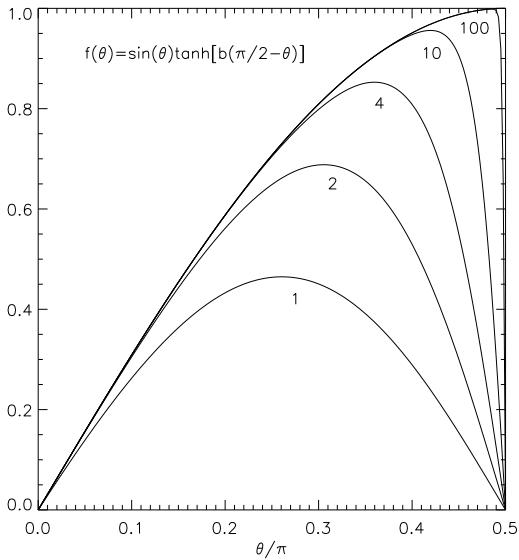}}}
\caption{
The latitude dependence of the toroidal field, as defined by 
Eq.(\ref{eq:tanh}) (only the field above the equator is displayed).
In this section the value $b=10$ is employed.
}
\label{fig:tanh}
\end{figure}

To understand what differences are to be expected, we have performed
also a series of simulations with a magnetic field given as
\be
B(r,\theta)=B_0 {r_0 \over r}\sin \theta \, \tanh [b(\pi/2-\theta)],
\label{eq:tanh}
\ee
where $B_0$ is the same parameter defined in Sect.~\ref{sec:model}. 
in terms of
$\sigma$, and $b$ is a free numerical parameter defining the width
of the striped wind region. In Fig.~\ref{fig:tanh} we show the angular
dependence of $B$ for various values of $b$: $b\to\infty$ corresponds
to the case discussed in all the previous sections, with 
$B \propto \sin\theta$, $b=10$ is
the case that will be considered in this section, while the field used 
in KL03 basically coincides with 
that in Eq.~(\ref{eq:tanh}) 
for $b\approx 0.7$.

\begin{figure*}
\centerline{
\resizebox{1.1\hsize}{!}{
\includegraphics{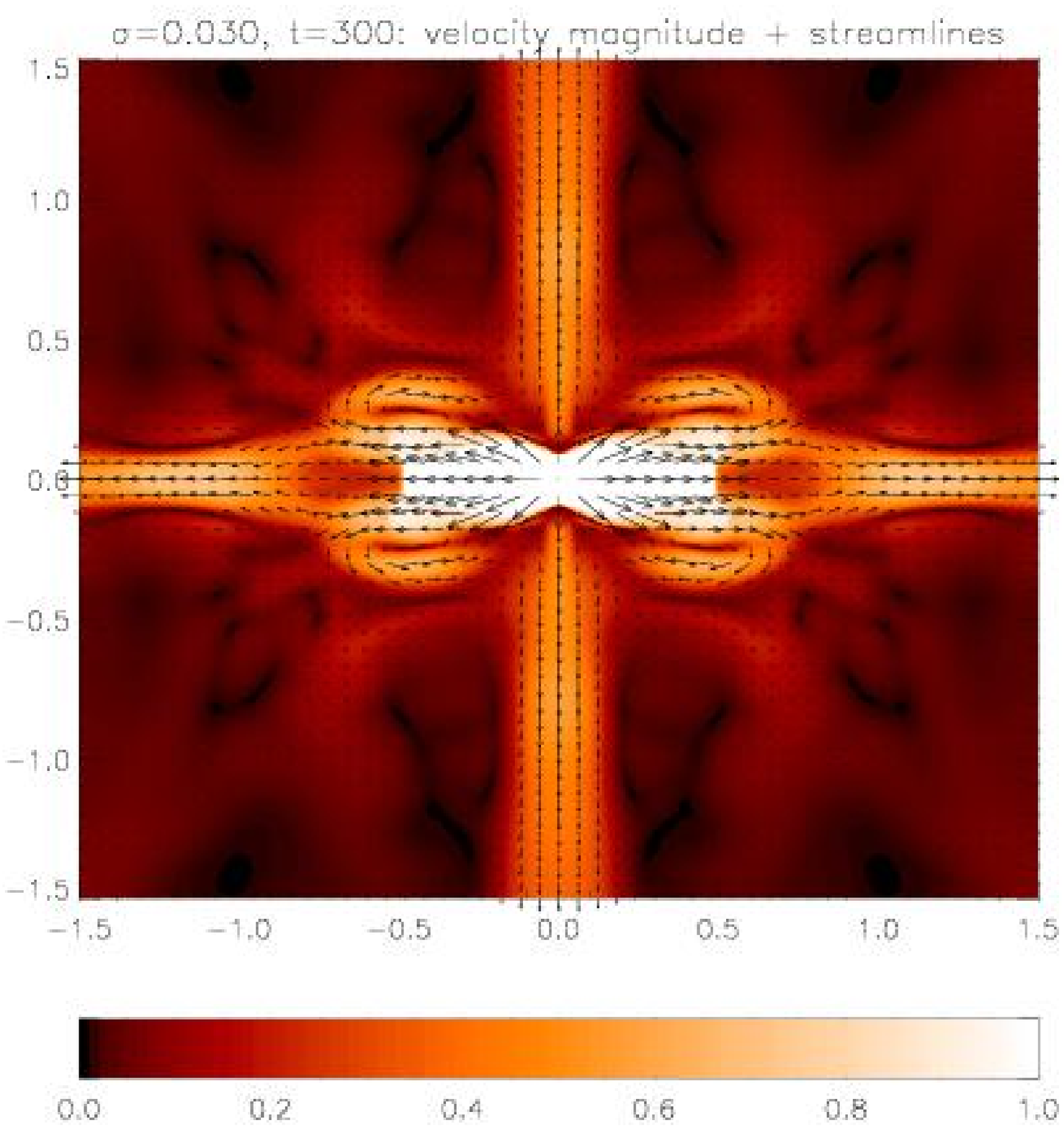}
\includegraphics{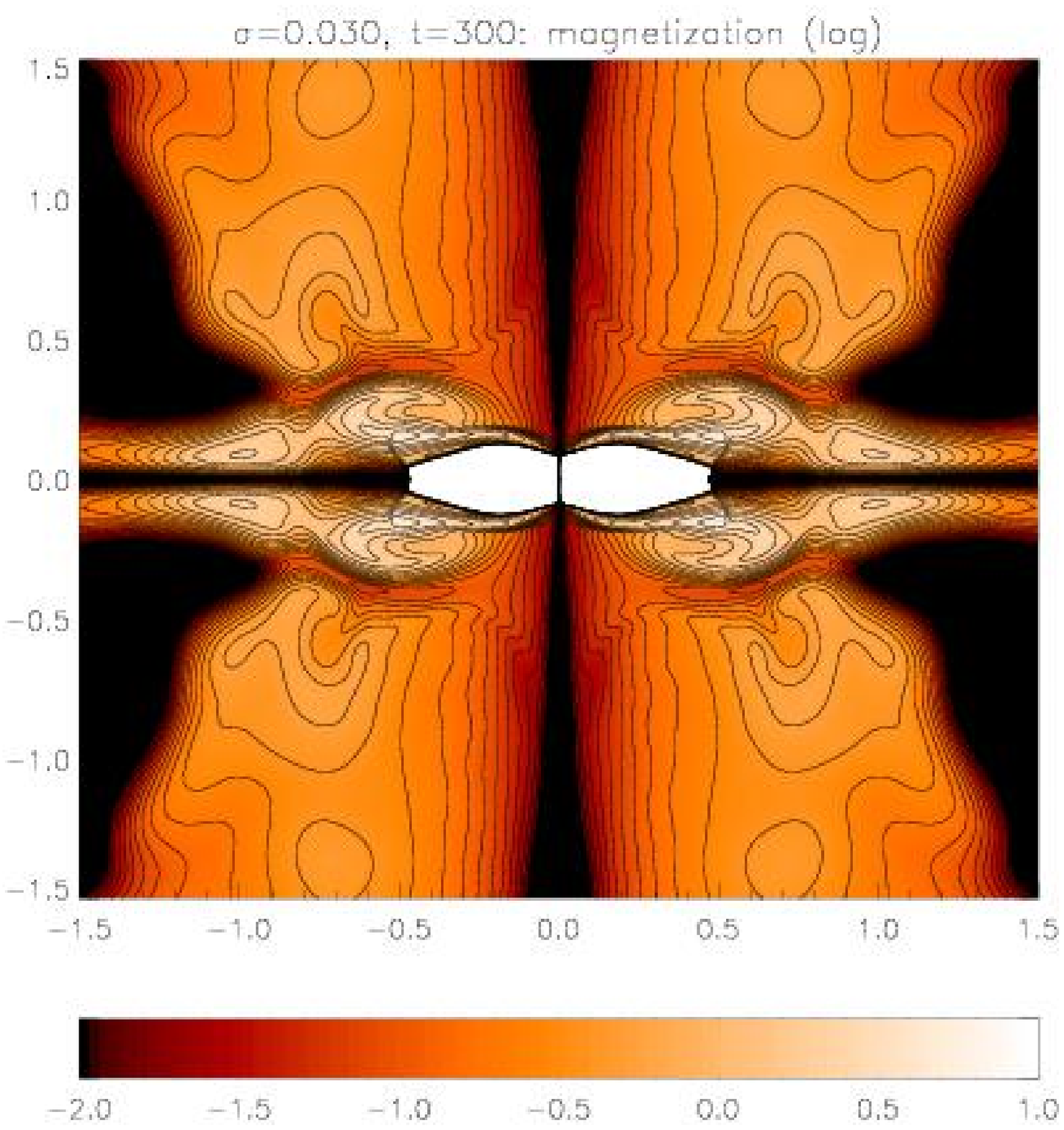}
\includegraphics{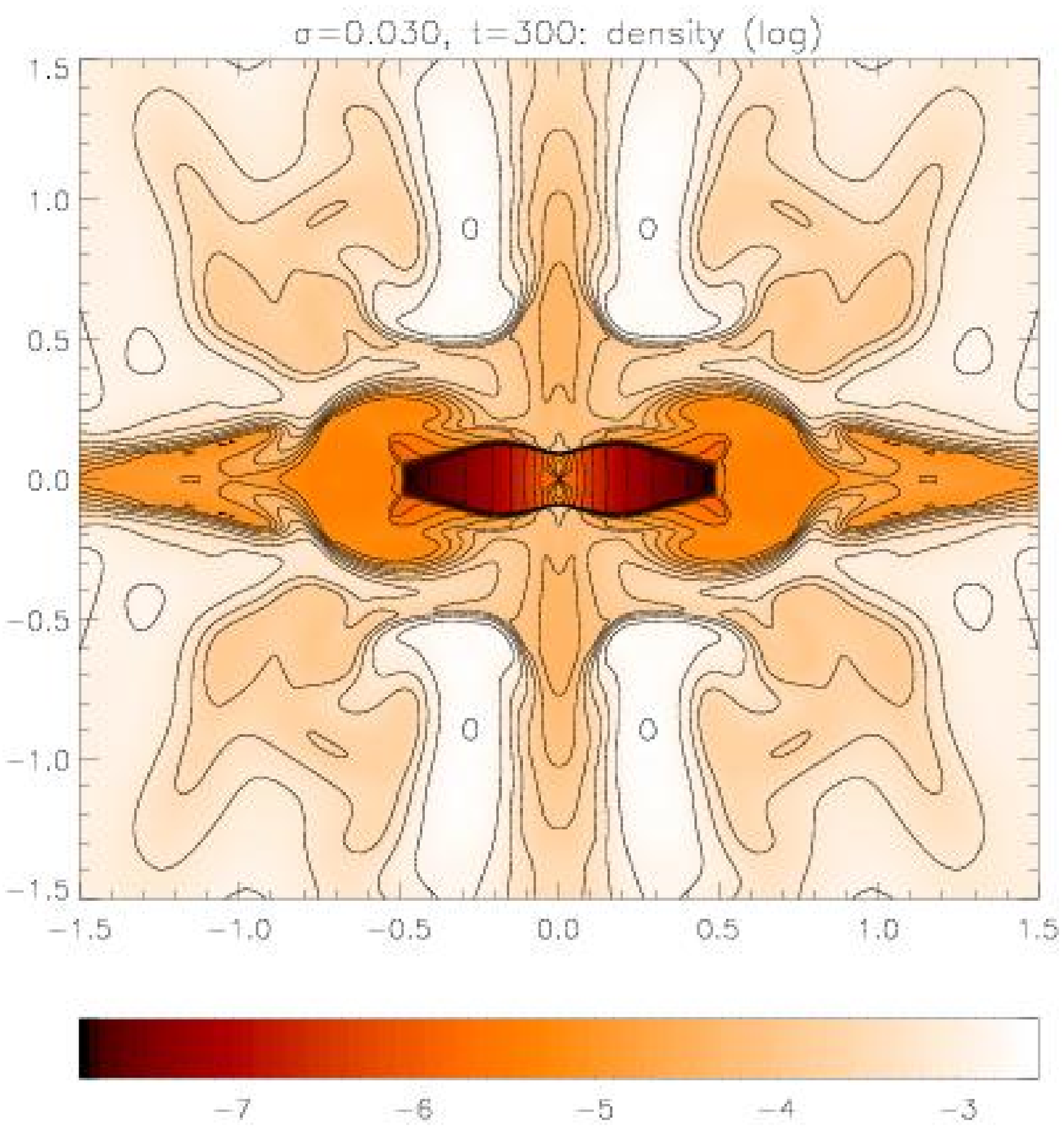}
}}
\caption{
Same quantities as in Fig.~\ref{fig:inner} at time $t=300$, for a case
with $\sigma=0.03$ but with the field shape in the wind
region given by Eq.~(\ref{eq:tanh}).
}
\label{fig:inner_tanh}
\end{figure*}

In Fig.~\ref{fig:inner_tanh} we show the same quantities as in
Fig.~\ref{fig:inner} at time $t=300$, again for the case $\sigma=0.03$ 
but using the magnetic field of Eq.~(\ref{eq:tanh}) with $b=10$.
The presence of an unmagnetized region around the equator (where the
magnetic field was the highest in the previous cases) even as
narrow as in the present case, is sufficient to change the picture
completely. Now the TS is able to move further out at the equator. 
Around the narrow channel to which the equatorial outflow is confined,
the magnetization grows rapidly and the hoop stresses are
effective. As a consequence, collimation of a polar outflow still 
occurs, although the fraction of plasma now involved is less than 
when the magnetic field was non-vanishing at the equator.
In particular (see Fig.~\ref{fig:ts_struct} for comparison) the high
speed funnel, instead of being completely diverted to the axis,
splits into an equatorial outflow and in a backflow toward the axis.
The corresponding streamlines now lie very close to the TS, and
correspondingly also the polar jet forms much closer to the origin.

\section{Conclusions}
\label{sec:final}

In this paper the structure and evolution of PWNe interacting with
the surrounding SN ejecta has been analyzed by means of relativistic MHD
axisymmetric simulations. Our main goal has been here the investigation 
of the mechanism originating the polar jets which are observed in X-rays 
in a growing number of PWNe.

The most recent and promising analytical studies (Bogovalov \& Khangoulian 
\cite{bk2}, Lyubarsky \cite{lyu02}) 
start with the assumption that the pulsar wind is highly anisotropic,
with a much larger energy flux at the equator than at the pole.
Such a wind, interacting with the expanding SN ejecta produces a hot 
magnetized bubble with a torus-jet structure, as observed. The polar jet 
is originated by the magnetic hoop stresses in the PWN that, 
for high enough values of the wind magnetization parameter $\sigma$,
divert part of the equatorial flow toward the axis, collimating and
accelerating it.
The first numerical RMHD simulations (Amato et al. \cite{amato}; 
KL03) confirm this 
scenario, and the simulated synchrotron emission map in the latter cited work
strikingly resembles, at least qualitatively, the X-ray images of the 
Crab Nebula.

Here we have made an effort to improve on those preliminary simulations.
The equatorial relativistic wind has a Lorentz factor as high as $\gamma=100$,
the magnetization parameter $\sigma$ goes from $0.003$ up to $0.1$, which
leads to magnetic fields in the PWN well beyond equipartition. The
magnetic field shape assumed in the wind is that proper for an aligned
or weakly oblique rotator, while in KL03 
the assumed field is
far from the $\sin\theta$ dependence, with a very broad region of low
magnetization around the equator. The evolution is followed up to the 
beginning of the reverberation phase and comparison with previous 
models is made whenever possible.

The results show that the predicted self-similar evolution of the external
PWN boundary is well reproduced at low magnetizations, and before
reverberation effects begin. The expected elongation of the PWN due 
to magnetic pinching is also recovered, and we show how this effect 
increases in time and with $\sigma$. The elongation of the nebula 
appears to be independent on the wind anisotropy, measured in our model 
by the parameter $\alpha$.

In the inner part of the PWN the termination shock assumes an oblate
shape, as expected for an anisotropic wind energy flux. An equatorial
supersonic flow is produced in a complex shock structure. At intermediate
latitudes, where the TS is highly oblique, the downstream flow is still 
supersonic. The plasma moves along the front gradually focusing toward
the equator. This pattern holds in hydrodynamical simulations as well, 
since it is due only to the wind flux anisotropy. The equatorial flow 
that is driven by this focusing mechanism is supersonic, with typical
velocities of $v\approx 0.5 - 0.7c$, consistent with the values
inferred for motion in the equatorial plane of the Crab Nebula 
(Hester et al. \cite{hester02}).

When the wind magnetization is high enough ($\sigma\gsim 0.01$ in our
simulations) the
magnetic hoop stresses in the equatorial plane are so strong as to
suppress completely this flow after a few termination shock radii.
The plasma is then completely diverted toward the axis, where the 
magnetic compression finally drives a supersonic polar outflow with
velocities that are once again in agreement with the values observed 
in the Crab and Vela PWNe ($v\approx 0.3 - 0.7c$, see
Hester et al. \cite{hester02}; Pavlov et al. \cite{pavlov03}).
At later times the interaction of the polar jets with the contact
discontinuity density jump may cause additional effects. The flow
circulates all the way along the CD from high latitudes down
to the equatorial plane. Here, these large scale vortexes drag 
some dense material from the ejecta toward the origin, with the 
effect of confining the equatorial outflow to the very inner parts 
of the nebula. Notice that the circulation is the opposite (from the
equator to the pole) in the hydro and low $\sigma$ cases. The value of
$\sigma$ that distinguishes the two regimes that we call highly
and lowly magnetizatized depends on the 
expansion velocity of the CD: for nebulae expanding at a larger rate
we expect the transition to occur for higher wind magnetizations. 

Finally, for the high $\sigma$ cases, the flow pattern strongly depends 
on the Poynting flux distribution in the wind. In particular, if in a 
narrow region around the equator the magnetic field vanishes, an additional 
vortex appears in the circulation pattern. The equatorial supersonic flow is 
never suppressed completely: it reaches the CD and then it circulates
back toward the axis. The polar jet is still present and drives a vortex
circulating at higher latitudes in the opposite direction than the former.

In conclusion, axisymmetric relativistic MHD simulations of the interaction 
of pulsar winds with expanding SNRs are able to reproduce at least
qualitatively most of the structures seen in X-ray images: the overall 
toroidal structure of the plerion, the supersonic motions in the 
equatorial plane and, if the wind magnetization is high enough, also the 
presence of polar jets with supersonic velocities. A more detailed
study, with a larger sampling of the parameter space would be necessary
for a quantitative comparison with the observations. An interesting 
perspective that the present study suggests is the inference of the 
wind magnetic field structure from direct comparison between simulated 
synchrotron maps and X-ray observations. Preliminary results are
encouraging, although we prefer to leave quantitative comparisons
as future work.

Some of the observed features are anyway impossible to reproduce within
the present axisymmetric RMHD framework. Emission structures like knots 
and sprites might well be related to non-ideal effects, like magnetic 
reconnection and dissipation, that are non-trivial to deal with. 
Remaining within the RMHD approximation, a full 3-D setting would allow to
deal with some non-axisymmetric instabilities that may play an important 
role in PWNe: let us mention as an example the kink 
instability, which is likely to be at the origin of the bending of the jet 
(observed in both the Crab and Vela nebulae).
We also leave the study of these important physical processes as future work.

\begin{acknowledgements}
The authors thank J. Arons, R. Bandiera, S. Komissarov, Y. Lyubarsky, 
F. Pacini, and M. Velli for fruitful discussions. This work was partly 
supported by MIUR under grants Cofin 2001 and Cofin 2002.
\end{acknowledgements}

\end{document}